\documentclass[aps,prd,twocolumn,superscriptaddress,amsfont,graphicx,nofootinbib,preprintnumbers]{revtex4}%
\usepackage{color,graphicx,epsfig}
\usepackage{ifpdf}
\usepackage{amsmath}
\usepackage{bm}
\usepackage{color}
\usepackage[english]{babel}
\usepackage{graphicx}%
\usepackage{amsfonts}%
\usepackage{amssymb}
\usepackage{braket}
\usepackage{hyperref}

\definecolor{nicered}{rgb}{0.7,0.1,0.1}
\definecolor{nicegreen}{rgb}{0.1,0.5,0.1}
\hypersetup{colorlinks,citecolor= nicegreen,linkcolor= nicered}

\newcommand{\slashed}{\slash \hspace{-0.19cm}}

\newcommand{\beq}{\begin{equation}}
\newcommand{\eeq}{\end{equation}}
\newcommand{\bea}{\begin{eqnarray}}
\newcommand{\eea}{\end{eqnarray}}

\begin{document}

\def\LjubljanaFMF{Faculty of Mathematics and Physics, University of Ljubljana,
 Jadranska 19, 1000 Ljubljana, Slovenia }
\def\LjubljanaIJS{Jo\v zef Stefan Institute, Jamova 39, 1000 Ljubljana, Slovenia}
\def\CERN{CERN, Theory Division, CH-1211 Geneva 23, Switzerland}
\def\Zurich{Physik-Institut, Universit\"at Zu\"rich, CH-8057 Zu\"rich, Switzerland}
\def\Sarajevo{Faculty of Science, University of Sarajevo, Zmaja od Bosne 33-35, 71000 Sarajevo, Bosnia and Herzegovina}

\title{
Confronting lepton flavor universality violation in B decays with high-$p_T$ tau lepton searches at LHC
}

\author{Darius A. Faroughy} 
\email[Electronic address:]{darius.faroughy@ijs.si} 
\affiliation{\LjubljanaIJS}

\author{Admir Greljo} 
\email[Electronic address:]{admir@physik.uzh.ch} 
\affiliation{\Zurich}
\affiliation{\Sarajevo}

\author{Jernej F.\ Kamenik} 
\email[Electronic address:]{jernej.kamenik@cern.ch} 
\affiliation{\LjubljanaIJS}
\affiliation{\LjubljanaFMF}

\preprint{ZU-TH-35/16}

\date{\today}
\begin{abstract}
{We confront the indications of lepton flavor universality (LFU) violation observed in semi-tauonic $B$ meson decays with new physics (NP) searches using high $p_T$ tau leptons at the LHC. Using effective field theory arguments we correlate possible non-standard contributions to semi-tauonic charged currents with the $\tau^+ \tau^-$ signature at high energy hadron colliders. Several representative standard model extensions  put forward to explain the anomaly are examined in detail: (i) weak triplet of color-neutral vector resonances, (ii) second Higgs doublet and (iii) scalar or (iv) vector leptoquark. We find that, in general, $\tau^+ \tau^-$ searches pose a serious challenge to NP explanations of the LFU anomaly. Recasting existing $8$~TeV and $13$~TeV LHC analyses, stringent limits are set on all considered {simplified} models.  Future projections of the $\tau^+ \tau^-$ constraints as well as {caveats in interpreting them within more elaborate models are also discussed.}
}

\end{abstract}

\maketitle

\section{Introduction}

Lepton flavor universality (LFU) of weak interactions is one of the key predictions of the standard model (SM). Experimentally it has been probed at the percent level precision both directly in W decays at LEP~\cite{Agashe:2014kda}, but also indirectly via precision measurements of pion, kaon, D meson and tau lepton decays (see~\cite{Amhis:2014hma} for a review). Over the past several years, there has been accumulating evidence for departures from LFU in (semi)tauonic decays of B mesons. In particular, Babar~\cite{Lees:2012xj, Lees:2013uzd}, Belle~\cite{Huschle:2015rga,Abdesselam:2016cgx} and LHCb~\cite{Aaij:2015yra} have all reported measurements of LFU ratios
\beq
R(D^{(*)}) \equiv \frac{\Gamma(B \to D^{(*)} \tau \nu)}{\Gamma(B \to D^{(*)} \ell \nu)}\,,
\eeq
where $\ell=e,\mu$, systematically larger than the corresponding very precise SM predictions~\cite{Kamenik:2008tj, Lattice:2015rga,Na:2015kha, Fajfer:2012vx}. A recent HFAG average of all current measurements~\cite{Amhis:2014hma}
\begin{subequations}
{\bea
R(D^{*}) &=& (1.25\pm0.07) \times R(D^{*})_{\rm{SM}}\,,\\
R(D) &=& (1.32\pm0.16) \times R(D)_{\rm{SM}}\,,
\eea}
\end{subequations}
puts the combined significance of these excesses at the 
{$4.0~\sigma$} level {(assuming $R(D)=R(D^*)$ the significance exceeds $4.4~\sigma$)}. Both $R(D^{(*)}) $ exhibit deviations of the same order and a good fit to current data prefers an approximately universal enhancement of $\sim 30\%$ in both observables over their SM values. This relatively large effect in charged current mediated weak processes calls for new physics (NP) contributions in $b\to c \tau \nu$ transitions~\cite{Fajfer:2012jt}. At the tree level, the possibilities are reduced to the exchange of a charged scalar ($H^+$)~\cite{Crivellin:2012ye, Celis:2012dk} or vector ($W'$)~\cite{Greljo:2015mma,Boucenna:2016wpr} bosons, or alternatively colored states carrying baryon and lepton numbers (leptoquarks)~\cite{Dorsner:2013tla, Sakaki:2013bfa, Bauer:2015knc,Barbieri:2015yvd}. Importantly, all possibilities imply new charged (and possibly colored) states with masses at or below the TeV and with significant couplings to the third generation SM fermions, making them potential targets for direct searches at the LHC. The aim of the present work is to elucidate and quantify the current and future sensitivity of the LHC high-$p_T$ experiments (ATLAS and CMS) to such NP. In particular we will show that quite generally NP relevant to the $R(D^{(*)})$ anomalies can be efficiently probed using high-$p_T$  tau pair production at the LHC.

The rest of the paper is structured as follows. In section~\ref{sec:II} we employ effective field theory (EFT) arguments to correlate NP contributions to $R(D^{(*)})$ with high-$p_T$ signatures involving tau leptons. We then examine explicit single mediator extensions of the SM which can be matched onto the EFT addressing the LFU anomaly in Sec.~\ref{sec:III}. The resulting constraints coming from existing $\tau^+ \tau^-$ searches by ATLAS and CMS are presented in Sec.~\ref{sec:IV}. Future experimental prospects as well as possible directions for model building in order to alleviate $\tau^+ \tau^-$ constraints are discussed in Sec.~\ref{sec:VI}. 

\section{Effective field theory}
\label{sec:II}

At sufficiently low energies, the exchange of new massive particles induces effects which can be fully captured by the appearance of local higher dimensional operators within an effective field theory description where the SM contains all the relevant degrees of freedom. The leading contributions appear at operator dimension six. While the effects in semileptonic B decays can without loss of generality be described in terms of effective operators respecting the QCD and QED gauge symmetries relevant below the electroweak breaking scale $v_{\rm EW}\simeq 246$~GeV, this is certainly not suitable for processes 	occurring at LHC energies.
To fully explore the possible high-$p_T$ signatures associated with effects in $R(D^{(*)})$, a set of semileptonic dimension six operators invariant under the full SM gauge symmetry is required. In the following we adopt the following complete basis~\cite{Freytsis:2015qca,Alonso:2015sja} 
\bea
\small
\mathcal L^{\rm eff} &\supset& c^{ijkl}_{QQLL} (\bar Q_i \gamma_\mu \sigma^a Q_j) (\bar L_k \gamma^\mu \sigma_a L_l)\nonumber\\
& + & c^{ijkl}_{QuLe} (\bar Q_i u^j_R) i \sigma^2 (\bar L_k \ell^l_R) 
 +  c^{ijkl}_{dQLe} (\bar d^i_R Q_j ) (\bar  L_k \ell^l_R ) \nonumber\\
& + & c_{dQLe'}^{ijkl}(\bar d^i_R \sigma_{\mu\nu} Q_j ) (\bar L_k  \sigma^{\mu\nu} \ell^l_R ) + \rm h.c.\,,
\label{eq:eft}
\eea
where $Q_i = ({V_{ji}^*} u^j_{L},d^i_L)^T$ and $L_i = ({U_{j i}^*} \nu^j,\ell^i_L)^T$ are the SM quark and lepton weak doublets in a basis which coincides with the mass-ordered mass-eigenbasis of down-like quarks ($d^i$) and charged leptons ($\ell^i$), $V(U)$ is the CKM (PMNS) flavor mixing matrix and $\sigma^a$ are the Pauli matrices acting on $SU(2)_L$ indices (suppressed). Note that we have omitted a fifth possible operator $(\bar Q \sigma_{\mu\nu} u^j_R) i \sigma^2 (\bar L \sigma^{\mu\nu}\ell^l_R)$, which can be shown to be redundant. 

First observation that can be made at this point is that in addition to charged current ($u^i \to d^j \ell^k \nu^l$) transitions, all operators predict the appearance of neutral quark and lepton currents ($u^i \bar u^j \to \ell^k \bar \ell^l $ and/or $d^i \bar d^j \to \ell^k \bar \ell^l $).  We note however that this would no longer be true in presence of additional light neutral fermions ($\nu_R$) which could mimic the missing energy signature of SM neutrinos in $B \to D^{(*)} \tau \nu$ decays. Additional operators can namely be constructed by the simultaneous substitution $\ell_R \leftrightarrow \nu_R$ and $u_R \leftrightarrow d_R$ in Eq.~\eqref{eq:eft}, plus the operator $(\bar d^i_R \gamma_\mu u_R^j ) (\bar \nu_R \gamma^\mu \ell_R^k)$ which can affect $R(D^{(*)})$~\cite{Fajfer:2012jt} but do not contribute to neutral currents involving charged leptons. In the EFT approach such contributions thus seem to be transparent to the tauonic high-$p_T$ probes discussed in the following. Consequently we do not include operators involving $\nu_R$ in our EFT discussion. In Sec.~\ref{sec:III} however, we use an explicit dynamical model to show that specific UV solutions of the $R(D^{(*)})$ puzzle involving $\nu_R$ can still be susceptible to our constraints.

To proceed further, we need to specify the flavor structure of the operators. We work with a particular choice of  flavor alignment (consistent with an $U(2)$ flavor symmetry acting on the first two generations of SM fermions), namely $c^{ijkl}_{QQLL}  \simeq c_{QQLL} \delta_{i3}\delta_{j3}\delta_{k3}\delta_{l3}$, $c^{ijkl}_{dQLe}  \simeq c_{dQLe} \delta_{i3}\delta_{j3}\delta_{k3}\delta_{l3}$, $c^{ijkl}_{dQLe'}  \simeq c_{dQLe'} \delta_{i3}\delta_{j3}\delta_{k3}\delta_{l3}$ which is motivated by (1) the requirement that the dominant effects appear in charged currents coupling to $b$-quarks and tau-leptons, and (2) stringent constraints on flavor changing neutral currents (FCNCs) (see Refs.~\cite{Fajfer:2012jt, Freytsis:2015qca, Greljo:2015mma} for more detailed discussion on this point). Small deviations from this limit, consistent with existing flavor constraints, would however not affect our conclusions. A common and crucial consequence of these flavor structures is that $b\to c$ quark currents always carry additional flavor suppression of the order $\sim |V_{cb}| \simeq 0.04$ compared to the dominant $b\to t$ (charged current) and $b\to b$, $t\to t$ (neutral current) transitions.

The flavor structure of $c_{QuLe}$ requires a separate discussion however. In the down-quark mass basis used in Eq.~\eqref{eq:eft}, the simplest choice ensuring dominant effects appear in $b\to c\tau\nu$ would be $c^{ijkl}_{QuLe}  \simeq c_{QuLe} \delta_{i3}\delta_{j2}\delta_{k3}\delta_{l3}$. However this flavor structure leads to  potentially dangerous $c\to u$ FCNCs, suppressed only by order of $\sim |V_{ub}|\simeq 0.004$ compared to the leading charged current effects. A safer choice with respect to flavor constraints would be to impose flavor alignement in  the mass basis of up-like quarks. In both cases the dominant induced neutral current is in the $ t \to c $ sector, while $c \to c$ is suppressed or completely absent. However, it has been shown previously~\cite{Freytsis:2015qca}, that non-zero $c_{QuLe}$ alone cannot accommodate both $R(D^{(*)})$ and be consistent with the measurements of the corresponding decay spectra. In the next section we provide the matching relations for suitable combinations of EFT operators within explicit NP models. It turns out that models addressing $R(D^{(*)})$ through  $c_{QuLe}$ contributions generically induce additional operators at low energies which do lead to sizeable $b \to b$ and/or $c\to c$ neutral current transitions.



We are now in a position to identify the relevant LHC signatures at high $p_T$. 
The main focus of this work is on $\tau^+ \tau^-$ production from heavy flavor annihilation in the colliding protons ($b\bar b \to \tau^+ \tau^-$ and $c\bar c \to \tau^+ \tau^-$). Even though it is suppressed by small heavy quark PDFs, this signature has been demonstrated previously to be extremely constraining for a particular explicit NP model addressing the $R(D^{(*)})$  anomaly~\cite{Greljo:2015mma}, owing in particular to the $\sim 1/|V_{cb}|$ enhancement of the relevant $b\bar b \to \tau^+ \tau^-$ neutral current process over the charged $b\to c \tau \nu$ transition, as dictated by flavor constraints.  As discussed above, in the EW preserving limit and in absence of cancelations (to be discussed later) a similar conclusion can be reached individually for all terms in Eq.~\eqref{eq:eft} except the one proportional to $c_{QuLe}$. Obviously, no such flavor enhancement is there for the related charged current mediated process of $\tau^+ \nu$ production from $\bar b c$ annihilation. The resulting constraints thus turn out not to be competitive. All other signatures involve at least three particles in the final state of the high energy collision and are thus expected to be phase-space suppressed.\footnote{Exceptions arise in case of on-shell QCD or EW pair production of new particles, which is not captured by the EFT in Eq.~\eqref{eq:eft} and which we discuss on explicit simplified model examples in Sec.~\ref{sec:III}.} 
As we demonstrate in the next section using explicit models, these conclusions hold generally even in presence of on-shell production of heavy NP mediators.  A notable exception are top quark decays, which do present an orthogonal sensitive high-$p_T$ probe, relevant especially for light mediator masses below the top quark mass~\cite{LFUt}. In the following we thus restrict our analysis to mediator masses above $\sim 200$~GeV.

\section{Models}
\label{sec:III}

The different chiral structures being probed by $R(D^{(*)})$ single out  a handful of simplified single mediator models~\cite{Freytsis:2015qca}. In the following we consider the representative cases, where we extend the SM by a single field transforming non-trivially under the SM gauge group. 

\begin{table}[!htb!]
\begin{center}
\scalebox{1.0}{
\begin{tabular}{ | c | c | c |  }
\hline 
 &  Color singlet & Color triplet  \\[0.2cm]  \hline
Scalar &  2HDM  & Scalar LQ \\ [0.2cm]
Vector & $W'$  & Vector LQ \\ [0.2cm] \hline

\end{tabular}
} 
\caption{\label{table:simplifiedmediators} 
A set of simplified models generating $b \to c \tau \nu$ transition at tree level, classified according to the mediator spin and color.
}
\end{center}
\end{table}

\begin{figure}[t]
\includegraphics[width=.45\hsize]{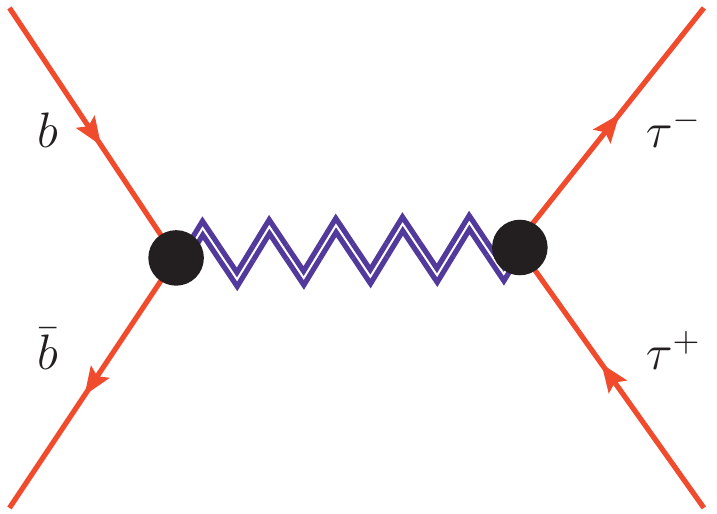}
\includegraphics[width=.45\hsize]{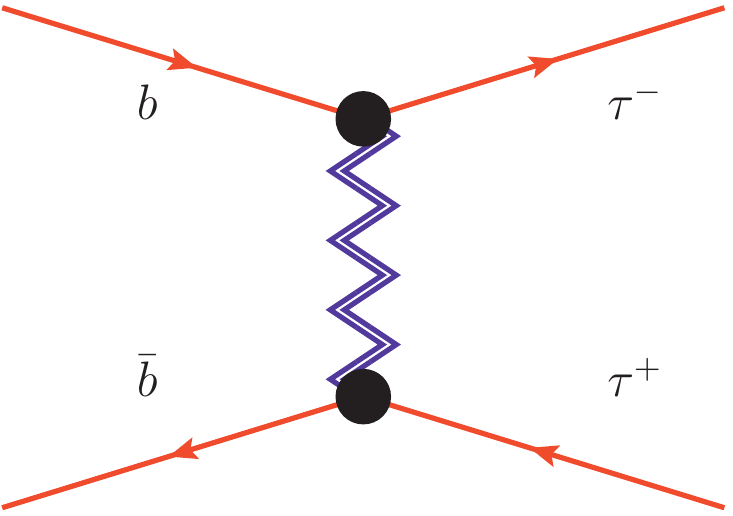}
\caption{Diagramatic representation of $s-$channel (left-hand side) and $t-$channel (right-hand side) resonance exhange (drawn in blue double see-saw lines) contributions to $b\bar b \to \tau^+ \tau^-$ process.}\label{fig:diag}
\end{figure}

{First categorization of single mediators is by color. While colorless intermediate states can only contribute to $b \to c \tau \nu$ transitions in the $s\equiv (p_b {-} p_c)^2$-channel, colored ones can be exchanged in the $t\equiv (p_b-p_\tau)^2$- or $u\equiv (p_b - p_\nu)^2$-channels. } {The colorless} fields thus need to appear in non-trivial $SU(2)_L$ multiplets {(doublets or triplets)} where the charged state mediating semileptonic charged currents is accompanied by one or more neutral states mediating neutral currents. Such models thus predict $\hat s\equiv(p_{\tau^+}+ p_{\tau^-})^2$-channel resonances in $\tau^+\tau^-$ production (see the left-hand side diagram in Fig.~\ref{fig:diag}). In addition to the relevant heavy quark and tau-lepton couplings, searches based on the on-shell production of these resonances depend crucially on the assumed width of the resonance, as we demonstrate below in Sec.~\ref{sec:IV}. Alternatively, {colored} mediators {(leptoquarks)} can be $SU(2)_L$ {singlets, doublets or triplets}, carrying baryon and lepton numbers. Consequently they will again mediate $\tau^+\tau^-$ production, this time through $\hat t\equiv(p_{b}- p_{\tau^-})^2$- or {$\hat u\equiv(p_{b}- p_{\tau^+})^2$-}channel exchange (see the right-hand side diagram in Fig.~\ref{fig:diag}). In this case a resonant enhancement of the high-$p_T$ signal is absent, however, the searches do not (crucially) depend on the assumed width (or equivalently possible other decay channels) of the mediators. In the following we examine the representative models for both cases {summarized in Table~\ref{table:simplifiedmediators}}.

\subsection{{ Vector triplet}}

A color-neutral real SU(2)$_L$ triplet of massive vectors $W'^a \sim W^{\prime \pm}, Z'$ can be coupled to the SM fermions via
\begin{align}
\mathcal L_{W'} &= -\frac{1}{4} W^{\prime a \mu\nu} W_{\mu\nu}^{\prime a} + \frac{M_{W'}^2}{2}W^{\prime a \mu} W_{\mu}^{\prime a} + W_{\mu}^{\prime a} J^{a \mu}_{W'}~\nonumber,\\
J^{a \mu}_{W'} &\equiv \lambda^q_{i j} \bar Q_i \gamma^\mu \sigma^a Q_j +  \lambda^\ell_{i j} \bar L_i \gamma^\mu \sigma^a L_j~.
\end{align}
Since the largest effects should involve $B$-mesons and tau leptons we assume $\lambda^{q(\ell)}_{i j}\simeq g^{}_{b(\tau)}  \delta_{i 3} \delta_{j 3}$, consistent with an $U(2)$ flavor symmetry~\cite{Greljo:2015mma}. {Departures from this limit in the quark sector are constrained by low energy flavor data, including meson mixing, rare $B$ decays, LFU and LFV in $\tau$ decays and neutrino physics, a detail analysis of which has been performed in Ref.~\cite{Greljo:2015mma}.\footnote{{Also, Ref.~\cite{Feruglio:2016gvd} considers leading RGE effects to correlate large NP contributions in $c_{QQLL}$ with observable LFU violations and FCNCs in the charged lepton sector. The resulting bounds can be (partially) relaxed in this model via direct tree level $W'$ contributions to the purely leptonic observables. }} The main implication is that the LHC phenomenology of heavy vectors is predominantly determined by their couplings to the third generation fermions ($g_b$ and $g_\tau$).  } 
The main constraint on $g_{b}$ comes from its contribution to CP violation in $D^0$ mixing yielding $g_{b}/M_{W'} < 2.2\,{\rm TeV}^{-1}$~\cite{Carrasco:2015pra}. On the other hand lepton flavor mixing effects induced by finite neutrino masses can be neglected and thus a single lepton flavor combination written above suffices without loss of generality. 
 
{In addition, electroweak precision data require $W'$ and $Z'$ components of $W^{\prime a}$ to be degenerate up to $\mathcal O(\%)$~\cite{Pappadopulo:2014qza}, with two important implications: (1) it allows to correlate NP in charged currents at low energies and neutral resonance searches at high-$p_T$; (2) the robust LEP bounds on pair production of charged bosons decaying to $\tau \nu$ final states~\cite{Abbiendi:2013hk} can be used to constrain the $Z'$ mass from below $M_{Z'} \simeq M_{W'} \gtrsim 100$\,GeV.
Finally, $W'^{a}$ coupling to the Higgs current ($W'_{a} H^\dagger \sigma^a \stackrel{\leftrightarrow}{D}_\mu H$) needs to be suppressed~\cite{Greljo:2015mma}, and thus irrelevant for the phenomenological discussions at LHC. 
 
Integrating out heavy $W'^a$ at tree level, generates the four-fermion operator,
 \beq
\mathcal L^{\rm eff}_{W'} = -\frac{1}{2 M_{W'}^2} J^{a \mu}_{W'} J^{a \mu}_{W'}~,
 \eeq
and after expanding $SU(2)_L$ indices,
{\footnotesize
\bea
\mathcal L^{\rm eff}_{W'} &\supset& - \frac{ \lambda^q_{i j} \lambda^{\ell}_{kl} }{M^2_{W'}}(\bar Q_i \gamma_\mu \sigma^a Q_j) (\bar L_k \gamma^\mu \sigma^a L_l)  \nonumber\\
&\supset& - \frac{g_b g_\tau}{M^2_{W'}} \left(2 V_{c b} \bar c_L \gamma^\mu b_L \bar \tau_L \gamma_\mu \nu_L + \bar b_L \gamma^\mu b_L \bar \tau_L \gamma_\mu \tau_L \right)~.
\eea} 
The resolution of the $R(D^{(*)})$ anomaly requires $c_{QQLL}^{}\equiv-g_b g_\tau / M^2_{W'} \simeq -(2.1 \pm 0.5)\,{\rm TeV}^{-2}$, leading at the same time to potentially large $b~ \bar b \to Z' \to \tau^+ \tau^-$ signal at the LHC.

Production and decay phenomenology of $W'$ and $Z'$ at the LHC have already been discussed in Refs.~\cite{Greljo:2015mma,Buttazzo:2016kid}, showing that the $R(D^*)$ anomaly cannot be addressed consistently in presence of a narrow $Z'$ decaying to $\tau^+\tau^-$. Here we significantly extend these previous works by recasting existing LHC $\tau \tau$ searches including possible large resonance width effects in order to properly extract the LHC limits on this model (see Section~\ref{sec:TripletBounds} for results). }
 
 \begin{figure}[t]
\includegraphics[width=0.6\hsize]{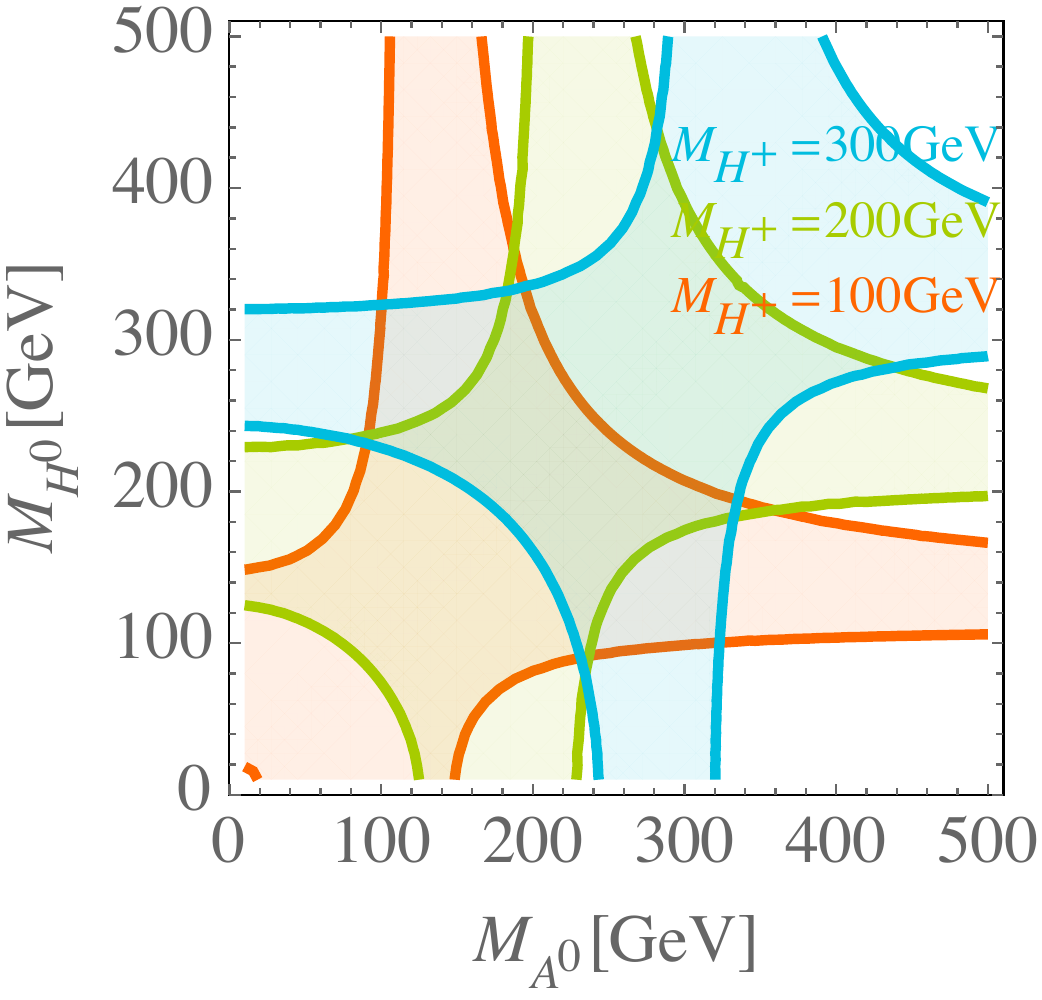}
\caption{Electroweak precision constraints on the masses of the scalars in the 2HDM, in the CP conserving and alignment (inert) limit. Allowed regions are shaded in orange, green and blue for reference charged scalar masses of $M_{H^+}=100$\,GeV, 200\,GeV and 300\,GeV, respectively (in sequence from bottom-left to top-right). For a given value of $M_{H^+}$ values of $M_{A,H}$ outside of the corresponding shaded region are excluded at the $3\sigma$ level.}\label{fig:EWPO}
\end{figure}

\subsection{{ Scalar doublet}} 

Color-neutral $SU(2)$ doublet of massive scalars with hypercharge $Y=1/2$, $H' \sim (H^{+}, (H^0 + i  A^0)/\sqrt{2})$ has the renormalizable Lagrangian of the form 
\begin{align}
\mathcal L_{H'} &= |D^\mu H'|^2 - M_{H'}^2 |H'|^2 - \lambda_{H'} |H'|^4 - \delta V(H',H) \nonumber\\
&- Y_b \bar Q_3 H' b_R - Y_c \bar Q_3 \tilde H' c_R - Y_\tau \bar L_3 H' \tau_R + \rm h.c.\,,
\end{align}
where 
$\tilde H'_{} = i \sigma^2 H^{\prime * }$\, and $\delta V(H',H)$ parametrizes additional terms in the scalar potential which lead to splitting of $A,H^0,H^+$ masses and to mixing of $H^0$ with the SM Higgs boson ($h$) away from the alignment (inert) limit. We discuss the relevance of these effects below. Additional couplings to fermions,  not required by $B$ decay data, are severely constrained by neutral meson oscilations and/or LFU measurements in the $\pi, K, D_{q}$ meson and $\tau$ lepton decays, and we do not consider them any further.  

The $H'$ model can account for both $R(D^{(*)})$ and the observed decay spectra~\cite{Freytsis:2015qca} through simultaneous non-vanishing contributions to $c_{dQLe} = Y_b Y^*_\tau / M^2_{H^+} \simeq {(50\pm14)}\,{\rm TeV}^{-2}$ and $c_{QuLe} = Y_c Y_\tau / M^2_{H^+} \simeq (-1.6\pm0.5)\,{\rm TeV}^{-2}$ (renormalized at the $b$-quark mass scale $\mu_R \simeq 4.2$~GeV) via the exchange of the charged $H'$ component ($H^+$). The corresponding high-$p_T$ signatures at the LHC are on the other hand driven by $b\bar b \to (H^0, A) \to \tau^+ \tau^-$ processes. 

As in the vector triplet case, robust mass bounds can only be set on the charged states, in particular $M_{H^+}\gtrsim 90$~GeV as required by direct searches at LEP~\cite{Abbiendi:2013hk}.
However, in a general two higgs doublet model (2HDM), the masses of $A,H^0,H^+$ are independent parameters and no common $M_{H'}$ scale can be defined. Consequently, the mass scale suppressing charged currents entering $R(D^{(*)})$ ($M_{H^+}$) could be significantly different from the masses of neutral scalars ($H^0,A$) to be probed in the $\tau^+\tau^-$ final state at the LHC.
However, the spectrum is also subject to electroweak precision constraints. In particular, the extra scalar states contribute to the gauge boson vacuum polarizations, parametrized by the Peskin-Takeuchi parameters S and T. Working in the CP conserving and alignment (inert) limits, we can employ the known results~\cite{Barbieri:2006bg} for the relevant 2HDM contributions. Comparing these to the recent Gfitter fit of electroweak precision data~\cite{Baak:2014ora} we obtain the constraints shown in Fig.~\ref{fig:EWPO}. 
We have checked that similar results are obtained even for moderate departures from the alignment (inert) limit, as allowed by current Higgs precision measurements. We observe that both $A,H^0$ cannot be simultaneously arbitrarily decoupled in mass from $H^+$.  In particular, we find that at least one neutral scalar has to lie within $\sim 100$\,GeV of the charged state. This level of uncertainty needs to be kept in mind when interpreting the constraints on this model derived in Section~\ref{sec:THDMBounds}.

\subsection{{Vector Leptoquark}} 

One can also extend the SM with a vector leptoquark weak singlet, $U_\mu \equiv ({\bf 3},{\bf 1},2/3)$,\footnote{Similar conclusions also apply for an $SU(2)_L$ triplet model.} coupled to the left-handed quark and lepton currents~\cite{Fajfer:2015ycq,Barbieri:2015yvd,Buttazzo:2016kid},
\bea
\mathcal L_U &=&  ~-\frac{1}{2} U^\dagger_{\mu\nu} U^{\mu \nu} +M_U^2 U^\dagger_\mu U^\mu+(J^\mu_{U} U_\mu+\rm{h.c.})~,\\
J^\mu_{U} &\equiv&  \beta_{i j}~ \bar Q_i \gamma^\mu L_j\,,
\eea
where again we restrict our discussion to $\beta_{i j} \simeq g_U \delta_{3 i} \delta_{3 j}$, {consistent with a $U(2)$ flavor symmetry~\cite{Barbieri:2015yvd}. Low energy flavor phenomenology of such models has been discussed in Refs.~\cite{Barbieri:2015yvd,Buttazzo:2016kid}, implying that the third generation fermion couplings dominate the phenomenological discussion also at the LHC. 

Unlike in the case of colorless mediators, QCD induced leptoquark pair production can lead to a large signal rate at the LHC, thus yielding robust constraints on the leptoquark mass $M_U$.  In the exact $U(2)$ flavor limit, $\mathcal{B}(U\to t \nu)=\mathcal{B}(U\to b \tau)=0.5$. {Revisiting the ATLAS search~\cite{Aad:2015caa} for QCD pair-produced third generation scalar leptoquark in the $t \bar t \nu \bar \nu$ channel, Ref.~\cite{Barbieri:2015yvd}, excludes $M_U < 770$~GeV.} {For large $\beta_{ij}$, limits from leptoquark pair production are even more stringent due to extra contributions  from diagrams with leptons in the $t-$channel~\cite{Dorsner:2014axa}. }
 
Integrating out the heavy $U_\mu$ field at the tree level, the following effective dimension six interaction is generated
 \beq
\mathcal L^{\rm eff}_{U} = -\frac{1}{ M_{U}^2} J^{\mu \dagger}_{U} J^{\mu}_{U} ~.
 \eeq
{Using Fierz identities to match the above expression onto the operator basis in Eq.~\eqref{eq:eft}, one finds
{\footnotesize
\beq
 \mathcal L^{\rm eff}_{U} = -\frac{\beta_{il}\beta_{k j}^\dagger}{2 M^2_U}  [ (\bar Q_i \gamma_\mu \sigma^a Q_j) (\bar L_k \gamma^\mu \sigma_a L_l)+(\bar Q_i \gamma_\mu Q_j) (\bar L_k \gamma^\mu L_l)]~,
\eeq}
which finally leads to
\beq
 \mathcal L^{\rm eff}_{U} \supset -\frac{|g_U|^2}{M^2_U} \left[V_{c b} (\bar c_L \gamma^\mu b_L) (\bar \tau_L \gamma_\mu \nu_L )+ (\bar b_L \gamma^\mu b_L) (\bar \tau_L \gamma_\mu \tau_L) \right]~.
\eeq}
 The fit to $R(D^{(*)})$ anomaly requires $|g_U|^2 / M^2_{U} \equiv 2 |c_{QQLL}^{}| \simeq (4.3 \pm 1.0)\,{\rm TeV}^{-2}$.
 As a consequence, sizeable $b ~ \bar b \to \tau^+ \tau^-$ signal at LHC is induced via t-channel vector LQ exchange. {A recast of existing $\tau^+ \tau^-$ searches in this model is presented in the Section~\ref{sec:LQBounds}.}

\subsection{{ Scalar Leptoquark}} 

{
Finally, we analyze a model recently proposed in Ref.~\cite{Becirevic:2016yqi}, in which the SM is supplemented  by a scalar leptoquark weak doublet, $\Delta \equiv ({\bf 3},{\bf 2},1/6)$ and a fermionic SM singlet ($\nu_R$),\footnote{The case of several $\nu_R$ is a trivial generalization which does not affect our main results.} with the following Yukawa interactions,
\beq
\mathcal{L}_\Delta \supset  Y_L^{ij} \bar d_i (i\sigma_2 \Delta^*)^\dagger L_j + Y_R^{i\nu} \bar Q_i \Delta \nu_R + \rm{h.c.}~.
\eeq
The mass of the fermionic singlet is assumed to be below the experimental resolution of the semi-tauonic $B$ decay measurements, such that the excess of events is explained via the LQ mediated contribution with $\nu_R$ in the final state. Following Ref.~\cite{Becirevic:2016yqi}, the $R(D^{(*)})$ anomaly can be accommodated provided the model parameters (evaluated at mass scale of the leptoquark $\mu_R \sim 0.5-1$~TeV) take values respecting
\beq
\left( \frac{Y_R^{b\nu }~ Y_L^{b\tau *}}{g_w^2} \right) \left(\frac{M_W}{M_\Delta} \right)^2 = 1.2\pm0.3,
\label{eq:YRYL}
\eeq
(see Fig.~[1] in~\cite{Becirevic:2016yqi}) where $g_w \simeq 0.65$ and $M_W\simeq 80$\,GeV are the SM weak gauge coupling and $W$ boson mass, respectively. Considering an exhaustive set of flavor constraints, Ref.~\cite{Becirevic:2016yqi} finds that $Y_{L}^{s \tau}$,  $Y_{L}^{s \mu}$ and $Y_{R}^{s\nu}$ are in general constrained to be small, and we therefore do not consider them in our subsequent analysis. 

The $\Delta^{(2/3)}$ component decays dominantly to $b \tau$ and $t \nu$, while $\Delta^{(1/3)}$ decays to the $b\nu$ final state. As in the vector leptoquark case, QCD pair production can again be used to obtain constraints on the leptoquark mass $M_\Delta$.  In particular, ATLAS~\cite{Aad:2015caa} excludes at 95\% CL pair-produced third-generation scalar leptoquarks decaying exclusively to $b \bar b \nu 
\bar \nu$ for $M_\Delta < 625$~GeV and $t \bar t \nu \bar \nu$ for $M_\Delta < 640$~GeV, respectively. In addition, CMS~\cite{CMS:2016hsa} excludes  at 95\% CL $M_\Delta < 900$~GeV scalar leptoquarks decaying exclusively to $\tau$ leptons and $b$ quarks. Consequently, relatively large couplings are required in order to accommodate the $R(D^{(*)})$ anomaly. For example, $M_\Delta=650$~GeV, implies $|Y_R^{b\nu} Y_L^{b\tau}| = 34\pm9$. Imposing a (conservative) perturbativity condition on all partial decay widths $\Gamma(\Delta \to q_i \ell_j)/M_{\Delta} \lesssim 1$, leads to $|Y_{L,R}^{ij}| \lesssim 7.1$. 

In this model the $R(D^{(*)})$ resolution involves a light $\nu_R$ and thus cannot be matched onto the SM EFT  in Eq.~\eqref{eq:eft}. Nonetheless, sizable $b \bar b \to \tau \tau$ production at LHC is generated via $t$-channel $\Delta$ exchange, and can effectively constrain $|Y_L^{b\tau}|$ (see Section~\ref{sec:LQBounds}). A restrictive enough bound in conjunction with Eq.~\eqref{eq:YRYL} can in turn drive the $Y_R^{b \nu}$ coupling into the non-perturbative regime.}
 

\section{Sensitivity of existing LHC searches}
\label{sec:IV}

{In the following, we perform a recast of  several experimental searches employing the $\tau^+$ $\tau^-$ signature at the LHC, to set limits on the EFT operators introduced in Eq.~\eqref{eq:eft} as well as on the corresponding simplified models described in the previous section as possible UV completions beyond the EFT. These constraints are compared to the preferred regions of parameter space accommodating the $R(D^{(*)})$ anomalies.  }

\subsection{Recast of $\tau \tau$ resonance searches}

\paragraph*{\bf{ATLAS (8 TeV, 19.5~fb$^{-1}$):}}

The ATLAS collaboration has performed a search for narrow resonances decaying to the $\tau^- \tau^+$ final state at $8$~TeV $pp$ collisions with $19.5-20.3$~fb$^{-1}$ of data~\cite{Aad:2015osa}. 
{The details of the analysis and our recast methods are described in the Appendix.} 
We rely on the official statistical analysis performed by the ATLAS collaboration.  In particular, the observed 95\% CL upper limits on the allowed signal yields in the final selection bins are obtained by rescaling the observed 95\% CL upper limits on the production cross-section for the Sequential SM (SSM) as reported in Fig.~8 of~\cite{Aad:2015osa}. The rescaling factors are the signal event yields reported in Table 4 of~\cite{Aad:2015osa} divided by the predicted cross-section in SSM from Fig.~8 of~\cite{Aad:2015osa}. In particular, for the final selection bins defined with $m_T^{\rm{tot}}>$ 400, 500, 600, 750 and 850 GeV, the excluded number of signal events at 95\%  CL are $N_{\rm{evs}}>$ 21, 11, 5.3, 3.3 and 3.4, respectively.  Here the total transverse mass $m_T^{\text{tot}}$ of the visible part of $\tau_{\text{had}}\tau_{\text{had}}$ is defined by
\begin{equation}
\small
m_T^{\text{tot}}\equiv\sqrt{ m_T^2(\tau_1,\tau_2) + m_T^2(\slashed{E_T},\tau_1)+m_T^2(\slashed{E_T},\tau_2)}\,,
\label{eq:mT}
\end{equation}
where {\small$m_T(A,B)\!\!=\!\!\sqrt{p_T(A) p_T(B)[1-\cos\Delta\phi(A,B)]}$} is the transverse mass between objects $A$ and $B$, and $\slashed{ E_T}$ is the total missing transverse energy reconstructed in the event.
{As discussed in the Appendix, we perform (for each model) a montecarlo simulation of the $m^{\rm{tot}}_{T}$ distribution at the reconstruction level} in order to find the expected number of signal events in these bins. The point in the parameter space of a model is excluded if any of the above limits are exceeded. 

\paragraph*{\bf{ATLAS (13 TeV, 3.2~fb$^{-1}$):}}

The ATLAS collaboration has also performed a search for $\tau \tau$ resonances at $13$~TeV using 3.2 fb$^{-1}$ of data~\cite{atlas-13}. {We recast~\cite{atlas-13} by reproducing correctly the SM backgrounds, and injecting our signal (see Appendix for details). After performing the statistical analysis using the CLs method~\cite{Junk:1999kv} {on the $m_T^{\rm{tot}}$ distribution (Fig.~(4f) of Ref.~\cite{atlas-13}})}, we find that for the final selection bin defined via $m_T^{\rm{tot}}>$ 150, 186, 231, 287, 357, 444, 551 and 684 GeV, the excluded number of signal events at 95\%  CL are $N_{\rm{evs}}>$ 200, 190, 120, 50, 20, 9.2, 6.2 and 3.7, respectively.\footnote{Here we conservatively assume $\sim 10\%$ systematical uncertainty in the first four bins of Fig.~(4f) of Ref.~\cite{atlas-13}.}  Again, the point in a model's parameter space is excluded if the predicted number of events exceeds the limit in any of the bins. }


\paragraph*{\bf{ATLAS (13 TeV, 13.2~fb$^{-1}$):}} The ATLAS collaboration has recently released results on a search for the MSSM process $A^0/H^0\to\tau\tau$ using $13.2-13.3$~fb$^{-1}$ of collected data from $pp$-collisions at 13~TeV center-of-mass energy~\cite{atlas-MSSM-13}. We recast the search in the fully inclusive category described in the Appendix. We take advantage of the higher luminosity of this search and use it to probe models with $\tau\tau$ resonances in the lower mass region $200-700$~GeV which typically suffer from low sensitivity. {For this, we perform for each model a profile likelihood fit to a binned histogram distribution of $m_T^{\text{tot}}$ with seven bins bounded bellow by 150, 200, 250, 300, 350, 400, 450~GeV, respectively. To be as conservative as possible, we assume systematic uncertainties among bins reported in Figs.~(4d) and~(4e) of~\cite{atlas-MSSM-13} to be uncorrelated and add them linearly to obtain the inclusive ones}. Limits on the parameter space of a model are given at 95\% CL.

\subsection{Results}

We implemented the EFT operators as well as all the simplified models into {\tt Feynrules 2}~\cite{Alloul:2013bka} and generated $p p (b\bar b) \to \tau^+ \tau^-$ events using {\tt Madgraph 5}~\cite{Alwall:2014hca} at LO in QCD. The production cross-sections were then rescaled to the most precise known values in the literature (when available) for each specific case as described in detail below. The generated events were finally passed through the same simulation pipeline as described above and in the Appendix. 

\subsubsection{ {EFT exclusion limits}}

First, we demonstrate the LHC $\tau^+\tau^-$ search sensitivity within the EFT by switching on individual operators in Eq.~\eqref{eq:eft}. The respective production cross-sections are only known at LO in QCD and were computed using the NNPDF2.3~\cite{Ball:2012cx} PDF set at NLO in the 5-flavor scheme.
Comparing the predicted number of events after the final selection with the exclusions, we find at 95\%  CL :
\begin{subequations}
\bea
|c_{QQLL}^{}| &<& ~2.8~ (2.6)~{\rm TeV}^{-2}~\textrm{recast \cite{Aad:2015osa} (\cite{atlas-13})},\\
|c^{}_{dQLe}| &<&~2.1~ (1.9)~{\rm TeV}^{-2}~\textrm{recast \cite{Aad:2015osa} (\cite{atlas-13})},
\eea
\end{subequations}
while, as anticipated in Sec.~\ref{sec:II}, no relevant bound can be obtained on $c^{}_{QuLe}$.
For the scalar operator, 
which has a non-vanishing anomalous dimension and runs under the QCD RG evolution, we assume the representative renormalization scale to be within the highest $m_T^{\rm{tot}}$ bin, which dominates the experimental constraints -- $\mu_R \gtrsim 700$~GeV. Due to the very slow running of $\alpha_s$ above the top mass threshold, the associated ambiguity is expected to be small. 
On the other hand, these constraints should be taken with caution, since the LHC explores high $p_T$ momentum transfers where the EFT validity might break down. In the following, we thus rather derive more robust constraints on all explicit model examples introduced in Sec.~\ref{sec:III}. \\

\begin{figure}[t]
\includegraphics[width=1.\hsize, trim=70 0 10 0]{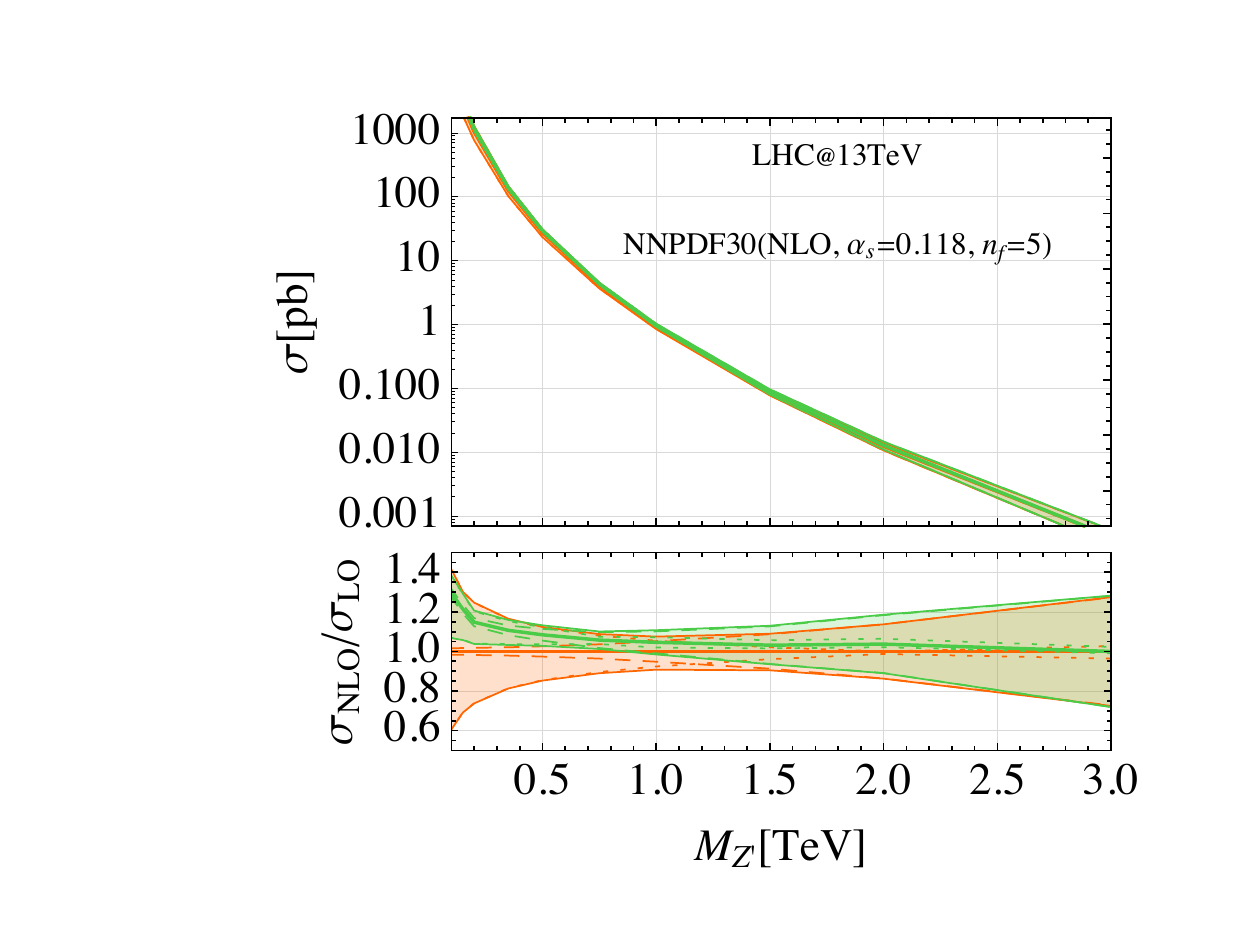}
\caption{Cross-sections for single on-shell $Z'$ production via bottom-bottom fusion at the 13~TeV LHC. The predictions obtained in the 5-flavor scheme at LO and NLO in QCD are shown in green and red shaded bands, respectively. See text for details.}\label{fig:xsec}
\end{figure}

\begin{figure}[t]
\includegraphics[width=0.7\hsize]{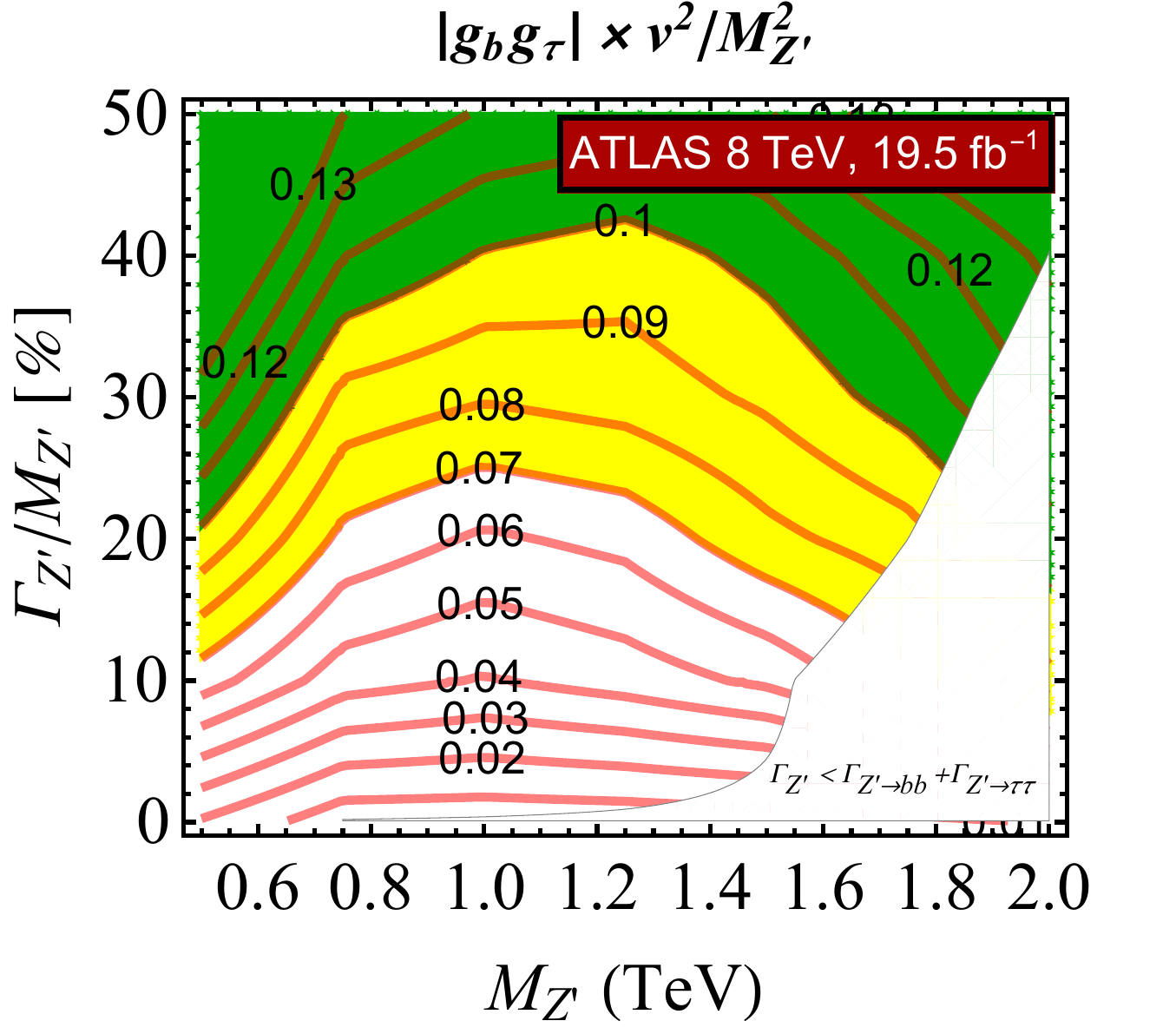}
\includegraphics[width=0.7\hsize]{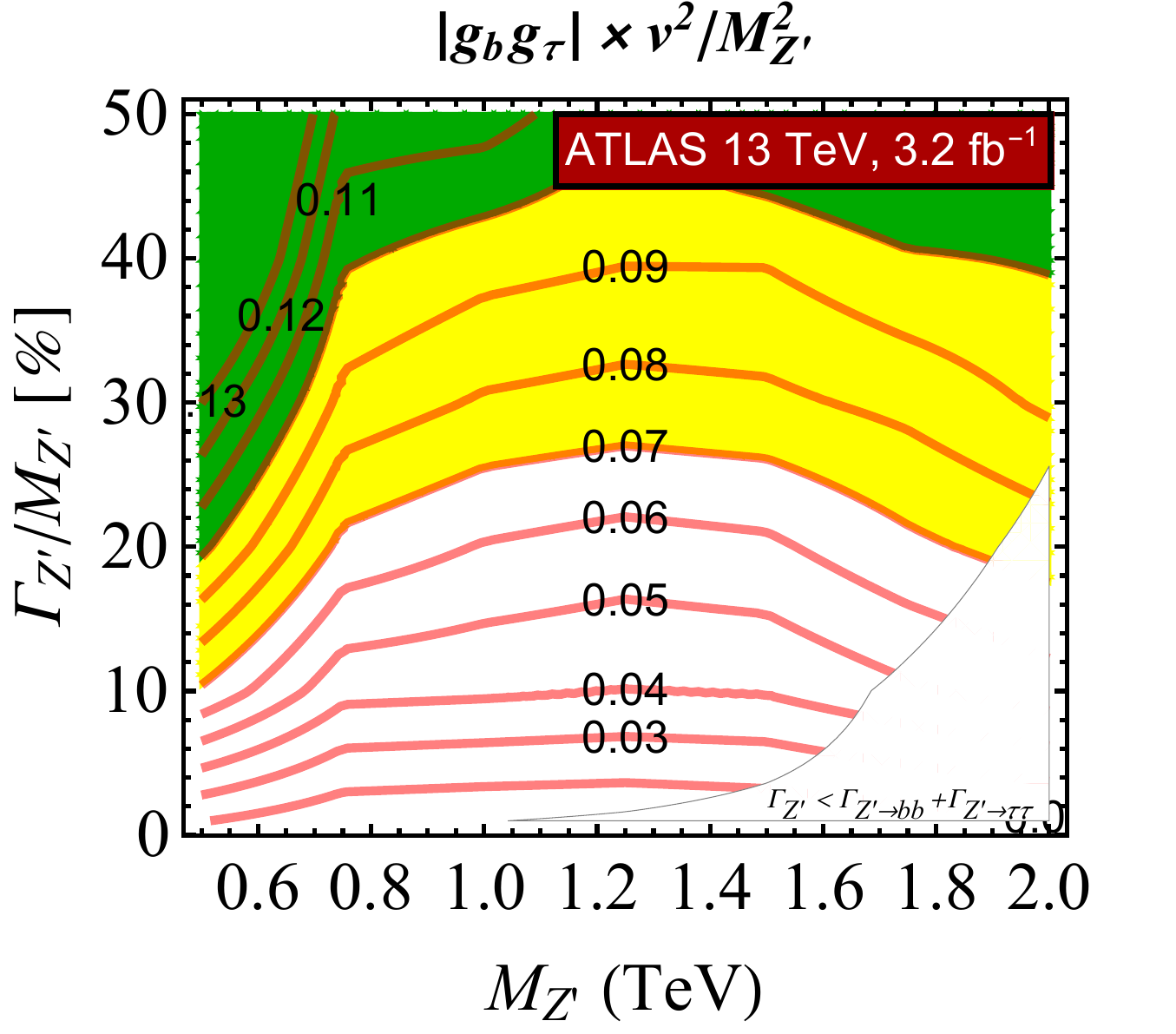}
\includegraphics[width=0.7\hsize]{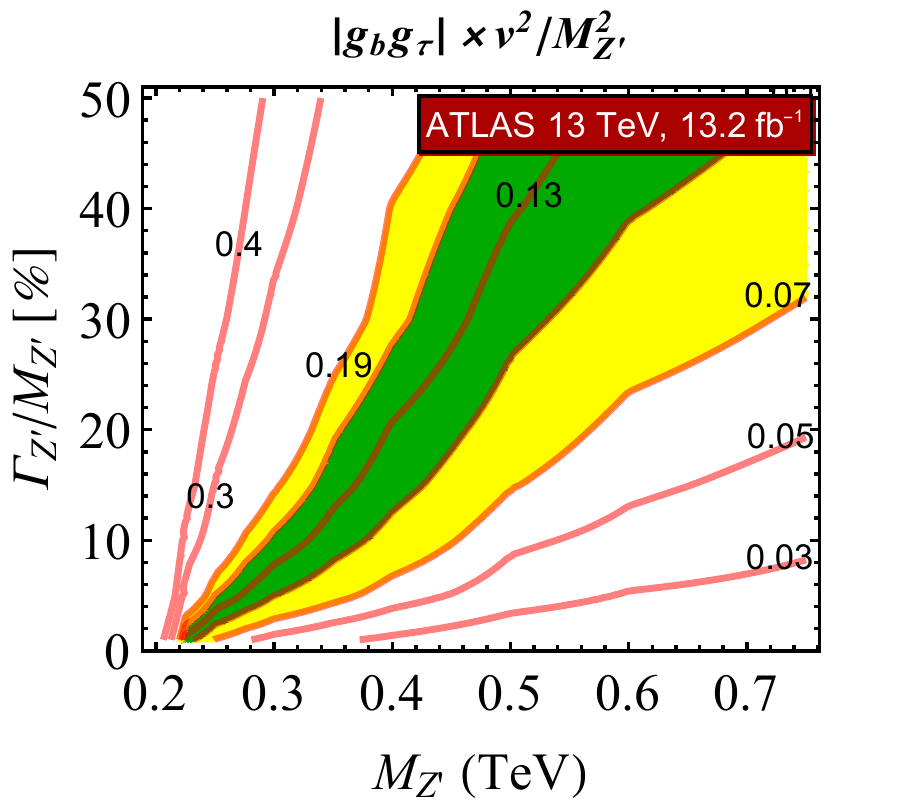}
\caption{Recast of ATLAS $\tau^+ \tau^-$ searches at 8~TeV~\cite{Aad:2015osa} (upper plot) 13~TeV with 3.2~fb${}^{-1}$~\cite{atlas-13} (middle plot) and 13~TeV with $13.2$~fb${}^{-1}$~\cite{atlas-MSSM-13} (lower plot) as exclusion limits on the $b \bar b$ induced spin-1 $\tau^+ \tau^-$ resonance ($b \bar b \to Z' \to \tau \tau$). Isolines shown in red represent upper limits on the combination $|g_b g_\tau|\times v^2/M_{Z'}^2$ as a function of the $Z'$ mass and total width. {The $R(D^{(*)})$  preferred regions $|g_b g_\tau|\times v^2/M_{Z'}^2 = (0.13\pm0.03)$ at $68\%$ and $95\%$ CL are shaded in green and yellow, respectively.}
}\label{fig:zpbounds}
\end{figure}


\subsubsection{ {Vector triplet exclusion limits}}
\label{sec:TripletBounds}

We start the discussion with a comment on the significance of the next-to-leading order (NLO) QCD corrections to $b \bar b$ induced $Z'$ production. In Fig.~\ref{fig:xsec} we plot the $Z'$ production cross-section at the 13~TeV LHC induced by $g_b=1$ as computed at the LO and NLO in QCD using aMC@NLO~\cite{Alwall:2014hca}, and shown in orange and green, respectively. We fixed the renormalization and factorization scales at $m_{Z'}$ and used the NNPDF3.0~\cite{Ball:2014uwa} set for PDFs in the NLO  5-flavor scheme. The perturbative (dotted contours), PDF (dashed contours) and total (shaded regions) uncertainties are also shown. The first are obtained independently varying factorisation and renormalisation scales within $\mu_F,\mu_R \in [0.5,2] M$, the second are given by the $68\%$ CL ranges when averaging over the PDF set. The total uncertainty is obtained by adding the perturbative and pdf uncertainties in quadrature. We observe that at low $Z'$ masses, perturbative uncertainty dominates, while above $\sim1$~TeV ($0.5$~TeV), the pdf uncertainty takes over at LO (NLO). Our numerical results and findings are consistent with those that have recently appeared in the literature for specific $Z'$ masses and SM-like couplings~\cite{Lim:2016wjo}. Similar results are found for 8TeV $pp$ colisions.
In setting bounds, we therefore rescale the LO simulation results to NLO production cross-section by applying the corresponding $K$-factor shown in Fig.~\ref{fig:xsec} (bottom) at the lower factorization, renormalization and $68\%$ CL PDF uncertainty ranges.

{The resulting $95\%$ CL upper limits on the $|g_b g_\tau|\times v^2/M_{Z'}^2$ for a given $Z'$ mass and total decay width, after recasting ATLAS 8 TeV~\cite{Aad:2015osa} (upper plot), 13 TeV with 3.2\,fb${}^{-1}$~\cite{atlas-13} (middle plot) and  13 TeV with $13.2$\,fb${}^{-1}$~\cite{atlas-MSSM-13} (lower plot)  $\tau^+\tau^-$ searches,  respectively,  are shown in Fig.~\ref{fig:zpbounds} and marked with red isolines. Note that this way of presenting  results is independent of the assumption on the existence of extra $Z'$ decay channels. The white region with gray border is not constrained since the assumed total width there is smaller than the minimum possible sum of the partial widths to $b \bar b$ and $\tau^+ \tau^-$ computed at the current experimental upper bound on $|g_b g_\tau|/M_{Z'}^2$. These exclusions are to be compared with the preferred value from the fit to the $R(D^{(*)})$ anomaly, $|g_b g_\tau|\times v^2/M_{Z'}^2 = (0.13\pm0.03)$, indicated in green (1$\sigma$) and yellow (2$\sigma$) shaded regions in the plot.


To conclude, for relatively heavy vectors $M_{W'}\gtrsim 500$~GeV within the vector triplet model, the resolution of the $R(D^{(*)})$ anomaly and consistency with existing $\tau^+ \tau^-$ resonance searches at the LHC require a very large $Z'$ total decay width. Perturbative calculations arguably fail in this regime. In other words, within the weakly coupled regime of this setup the resolution of the $R(D^{(*)})$ anomalies cannot be reconciled with existing LHC $\tau^+\tau^-$ searches.} On the other hand, interestingly, a light $Z'$ resonance with $M_{Z'} \lesssim 400$\,GeV, a relatively small width and couplings compatible with the $W'$ resolution of the $R(D^{(*)})$ anomaly is not excluded by our $\tau^+\tau^-$ search recast. Note, however, that our analysis is by no means optimized as we are forced to use a certain fixed number of bins and their sizes and cannot leverage the full control of experimental systematics. 

\begin{figure}[t]
\includegraphics[width=0.8\hsize]{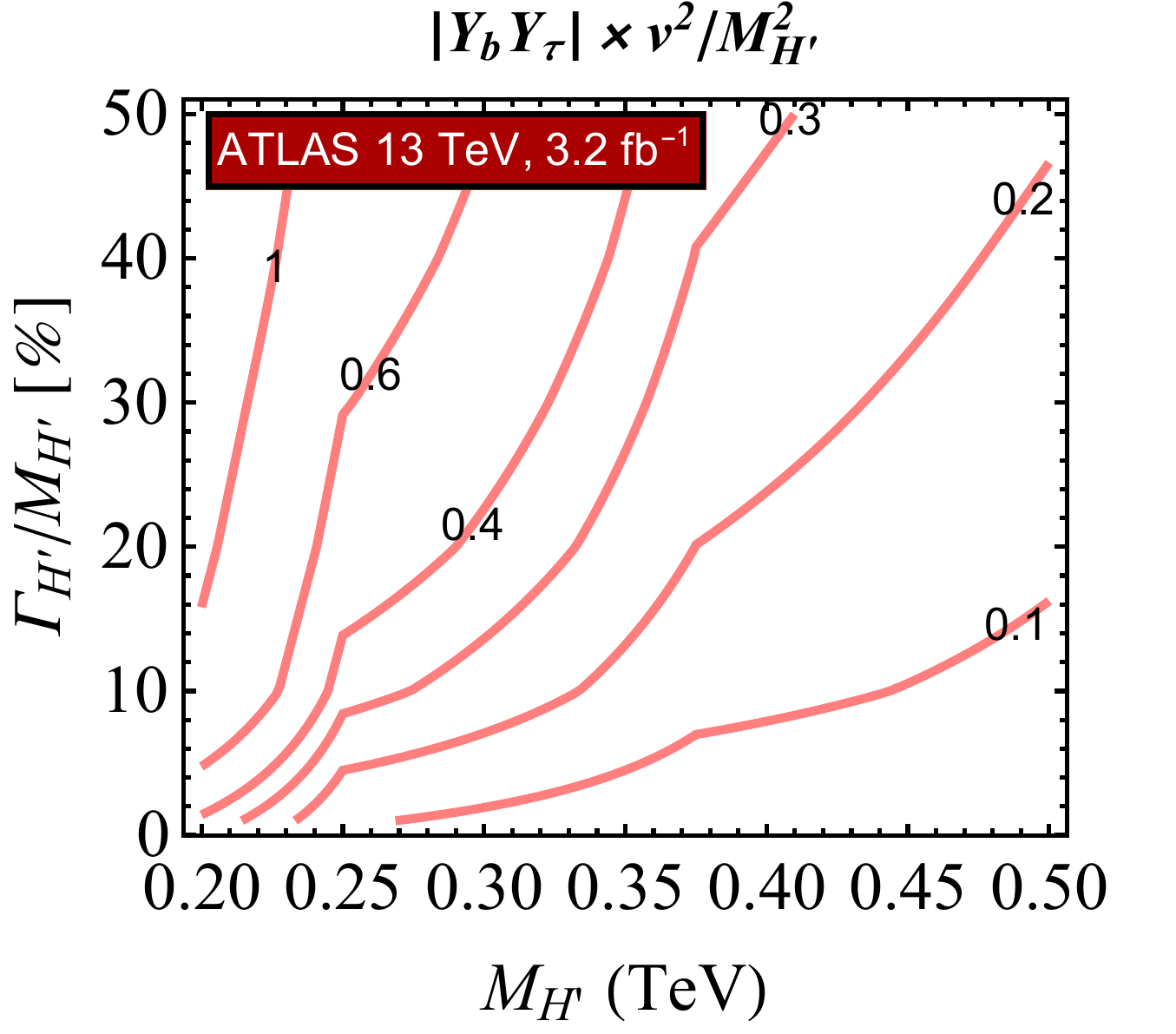}
\includegraphics[width=0.8\hsize]{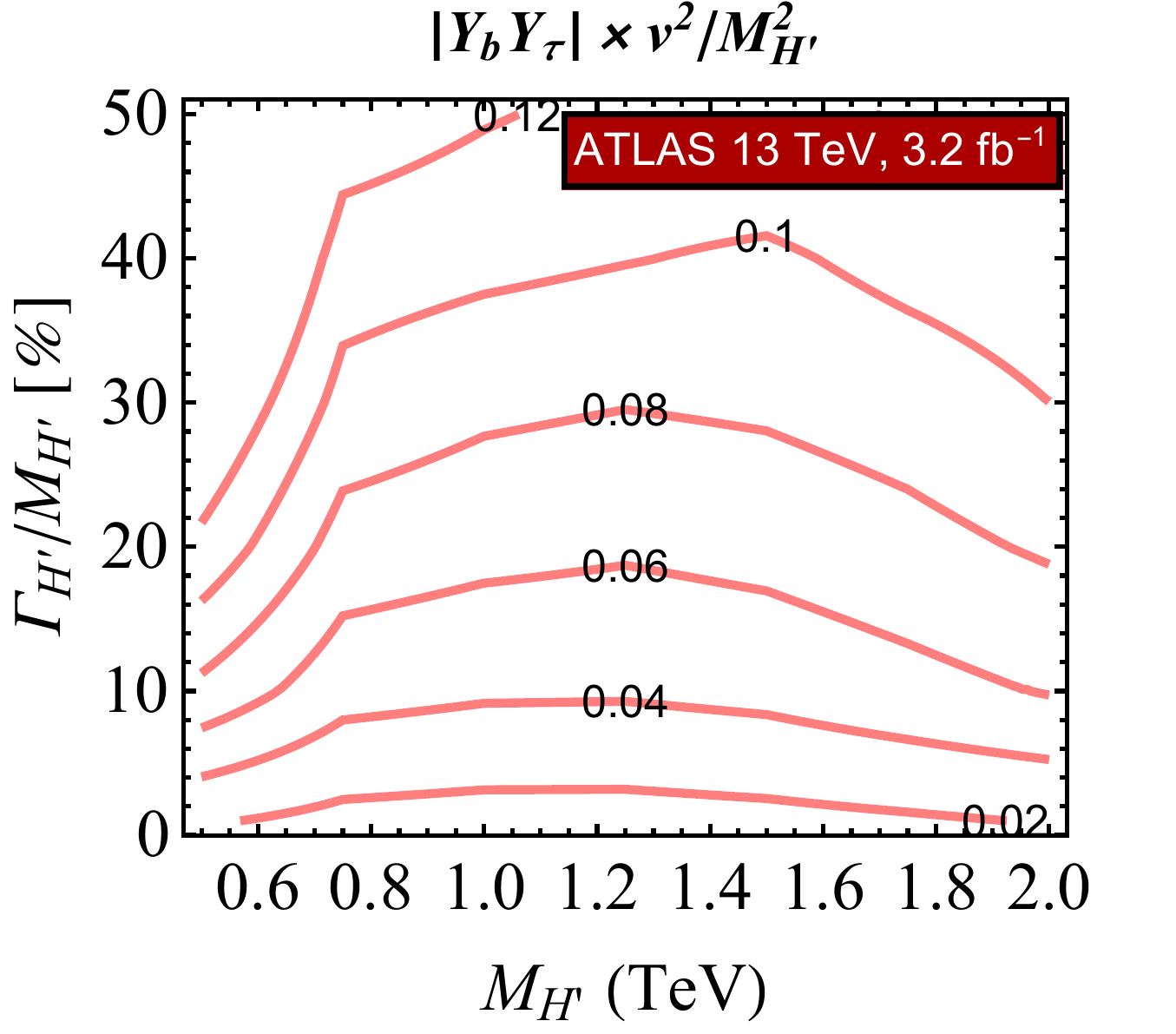}
\caption{ATLAS (13~TeV, $3.2$~fb$^{-1}$) $\tau \tau$ search~\cite{atlas-13} exclusion limits on $b \bar b \to H^0 \to \tau \tau$ resonances. {The preferred value from the fit to the $R(D^{(*)})$ anomaly is $Y_b Y_\tau^* \times v^2 / M_{H^+}^2 = (2.9\pm0.8)$.} }\label{fig:2HDM}
\end{figure}

\subsubsection{{ 2HDM exclusion limits}}
\label{sec:THDMBounds}

The cross-sections for $A, H^0$ production from $b\bar b$ annihilation can be estimated at NNLO in QCD using the Higgs cross-section WG results~\cite{Dittmaier:2011ti}. While the results are directly applicable for the CP even state $H^0$, 
we expect them to hold as a good approximation also for a heavy CP-odd $A^0$ due to the restoration of chiral symmetry when $m_{b}/m_{H'} \ll 1$\,. We have checked explicitly that differences between scalar and pseudoscalar production are negligible up to NLO ~\cite{Drees:1990dq} for the interesting mass region $m_{A^0,H^0}\gtrsim 200$~GeV. In setting bounds, we therefore rescale the LO simulation results to the Higgs cross-section WG production cross-sections~\cite{Dittmaier:2011ti} taken at the lower factorization, renormalization and $68\%$ CL PDF uncertainty ranges.



Conservatively considering only a single neutral scalar resonance contribution (denoted by $H'$ meaning either $A^0$ or $H^0$), we show the resulting $95\%$ CL upper limits on the $|Y_b Y_\tau|\times v^2/M_{H'}^2$ (evaluated at the $b$-quark mass scale $\mu_R \simeq 4.3$~GeV) after recasting the ATLAS 13 TeV~\cite{atlas-13}  $\tau^+\tau^-$ search in Fig.~\ref{fig:2HDM}. 
We observe that even after accounting for the possible $\mathcal O(100\,{\rm GeV})$ mass splitting between the charged and the lightest neutral state within the scalar $H'$ doublet, the $R(D^{(*)})$ preferred value $Y_b Y_\tau^* \times v^2 / M_{H^+}^2 = (2.9\pm0.8)$ cannot be reconciled with existing $\tau^+ \tau^-$ resonance searches at the LHC in the $m_{A,H^0}\gtrsim 200$~GeV region.\footnote{In case of $H'=H^0$ (with $A^0$ decoupled), small departures from the 2HDM alignment limit (i.e. non-zero $h-H^0$ mixing), consistent with existing experimental constraints, in particular on $h\to \tau^+\tau^-, b\bar b$~\cite{Khachatryan:2016vau} (see e.g.~\cite{Efrati:2016uuy}), can further mildly alleviate the bound due to somewhat reduced effective $Y_{b,\tau}$ couplings of $H^0$ compared to those of $H^+$. The required order of magnitude reduction can however not be achieved.}

\begin{figure}[t]
\includegraphics[width=0.8\hsize]{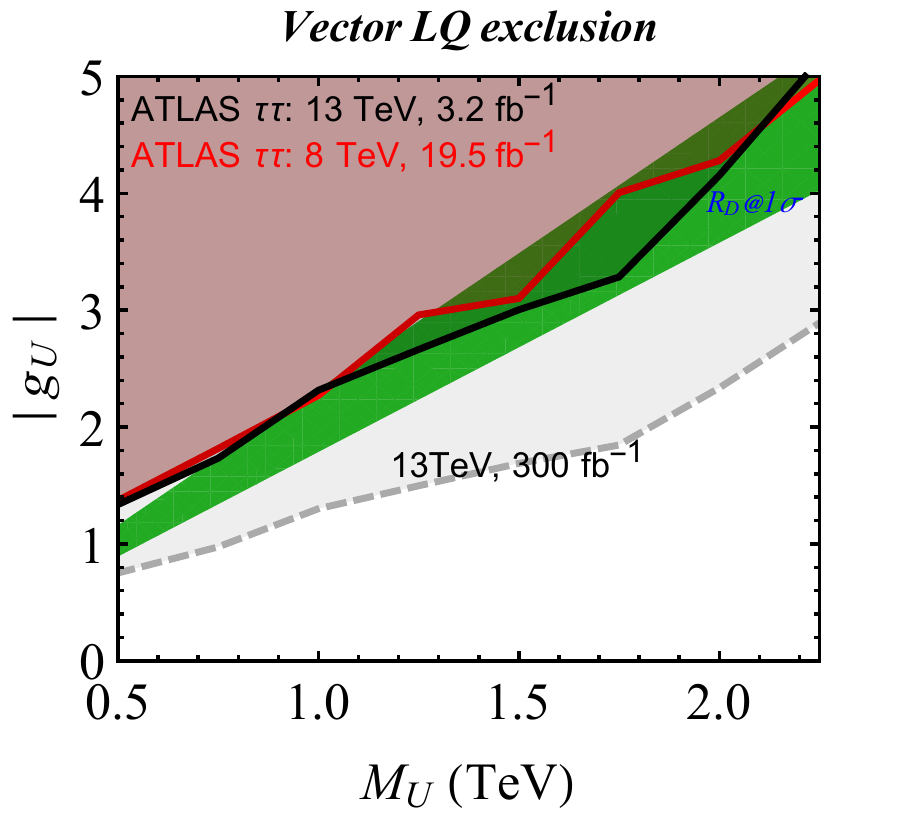}\\
\includegraphics[width=0.8\hsize]{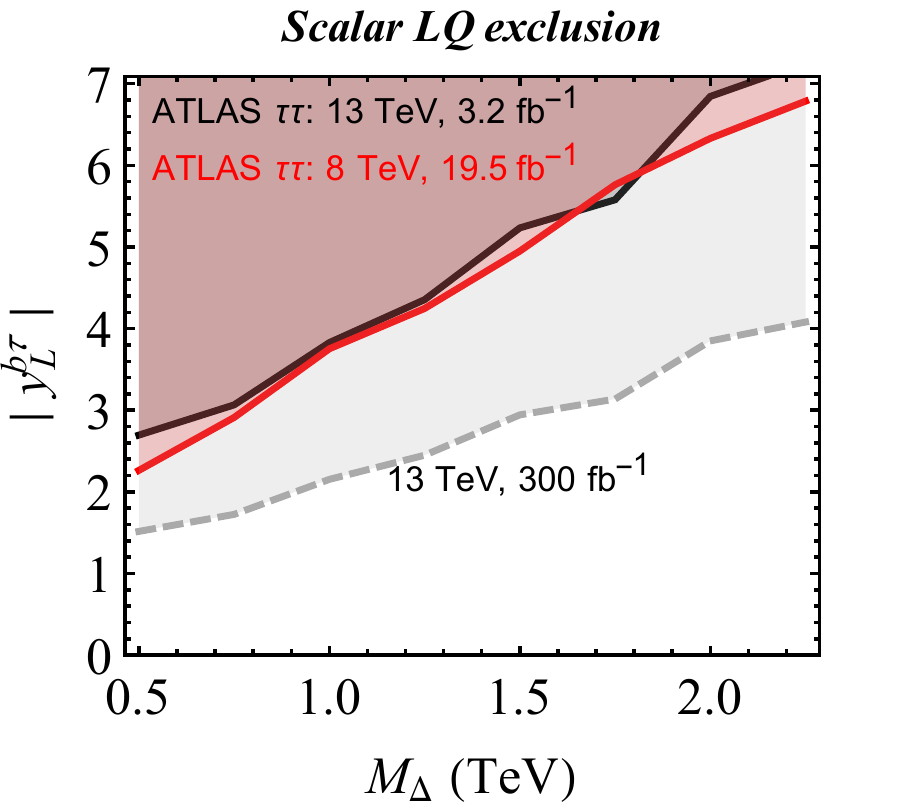}
\caption{(Upper plot) 8 TeV~\cite{Aad:2015osa} (13 TeV~\cite{atlas-13})  ATLAS $\tau^+ \tau^-$ search exclusion limits are shown in red (black) and $R(D^{(*)})$ preferred region in green for the vector leptoquark model. Projected 13 TeV limits for 300~fb$^{-1}$ are shown in grey. (Lower plot) the same search exclusion limits for the scalar leptoquark model.}\label{fig:vlq}
\end{figure}

\subsubsection{{ Scalar and Vector LQ exclusion limits}}
\label{sec:LQBounds}

{The $\tau^+\tau^-$ production through t-channel leptoquark exchange is only known at LO in QCD and we simulate it using the NNPDF2.3~\cite{Ball:2012cx} PDF set at NLO in the 5-flavor scheme. 
The exclusion limits for the vector leptoquark model from the recast of 8 TeV~\cite{Aad:2015osa} and 13 TeV~\cite{atlas-13} searches are shown in Fig.~\ref{fig:vlq} (top) in red and black shades, respectively. On the other hand, the preferred region at 68\% CL from $R(D^{(*)})$ anomaly is shown in green. In addition, projected exclusion limits at 13~TeV, with $300$~fb$^{-1}$ (assuming the present 13 TeV limits on the cross-section to scale with the square root of the luminosity ratio) are shown in gray. In this model, the $R(D^{(*)})$ anomaly explanation is already in some tension with existing $\tau^+\tau^-$ searches, and  future LHC Run-II data should resolve the issue conclusively. 

On the other hand, exclusion limits for the scalar leptoquark model are shown in Fig.~\ref{fig:vlq} (bottom). 
Although bounds can only be set on one of the two relevant couplings ($Y_L^{b\tau}$), we note that in order to keep $Y_R^{b \tau} Y_L^{b\tau}$ large enough to fully accommodate the $R(D^{(*)})$ anomaly (see Eq.~\eqref{eq:YRYL}), $Y_R^{b \tau}$ is pushed to non-perturbative values. }

\section{Conclusions and future prospects}
\label{sec:VI}

{
In this work we have discussed possible new dynamics that could explain the recent hints of LFU violation in semi-tauonic $B$ decays, and, in particular, {\it the physics case} for associated high $p_T$ searches at the LHC.

By employing effective field theory methods we have argued that in presence of non-standard effects in semi-leptonic charged currents, one in general expects signals also in neutral currents involving charged leptons. Moreover, requiring (i) dominant couplings to the third generation in order to explain the $R(D^{(*)})$ anomaly and (ii) protection from large FCNC in the down quark sector, neutral currents involving pure third generation fermions ($b b \to \tau \tau$) are $\sim 1/V_{cb}$ enhanced with respect to $b c \to \tau \nu$ charged currents, leading to potentially large signals at the LHC.

Indeed, by performing a recast of  existing $\tau^+ \tau^-$ resonance searches at the LHC, we set stringent limits on several representative simplified models involving:  a spin-1 colorless weak triplet (Sec.~\ref{sec:TripletBounds}), a 2HDM (Sec.~\ref{sec:THDMBounds}), a spin-0 or spin-1 leptoquark (Sec.~\ref{sec:LQBounds}). We find that in light of existing constraints it is paramount to consider (relatively) wide and (or) light resonances, and we encourage the experimental collaborations to perform and update their searches for $\tau^+ \tau^-$ resonances in a model independent way as illustrated in Figs.~\ref{fig:zpbounds}~and ~\ref{fig:2HDM}. At the same time, searches for non-resonant deviations in (the tails of) distributions are equally relevant as shown in the leptoquark analyses (see Fig.~\ref{fig:vlq}). 

Apart from the low mass region in the $W'$ and vector leptoquark models, all considered models are in tension with  existing $\tau^+ \tau^-$ LHC results. Near-future data is likely to cover all the remaining interesting parameter space for the vector leptoquark model. In the vector triplet model, electroweak pair production of light $W'$'s decaying to $\tau\nu$ could also provide competitive constraints at the LHC -- a detailed study is left for future work.

Possibilities within more elaborate NP models to avoid the current stringent constraints include (i) splitting the neutral and charged states in the weak multiplet or (ii) providing additional negatively interfering contributions in $\tau^+ \tau^-$ production, both of which require a degree of fine tuning. 


On the other hand, we note that the leptoquark singlet model proposed in Ref.~\cite{Bauer:2015knc} (see also~\cite{Dumont:2016xpj}) avoids existing LHC $\tau^+ \tau^-$ constraints due to the absence (suppression) of $b \bar b \to \tau^+ \tau^-$ ($c \bar c \to \tau^+ \tau^-$) processes, respectively. Third generation leptoquark searches (in particular, from QCD pair production~\cite{Khachatryan:2015bsa}) remain the best strategy in this case. Other possible signatures of single or associated leptoquark production in this model, e.g. ~monotops~\cite{Andrea:2011ws} could be interesting targets for future studies.}

\begin{acknowledgments}

We would like to thank Svjetlana Fajfer, Nejc Ko\v snik and Ilja Dor\v sner for useful discussions, and Gino Isidori for carefully reading the manuscript.  {AG thanks the Jo\v zef Stefan Institute (IJS) and Mainz Institute for Theoretical Physics (MITP) for hospitality while this work was being completed. AG is supported in part by the Swiss National Science Foundation (SNF) under contract 200021-159720. JFK acknowledges the financial support from the Slovenian Research Agency (research core funding No. P1-0035).  \\}

\end{acknowledgments}

\begin{appendix}
\section*{{Appendix}}

The exclusion limits presented in Sec.~\ref{sec:IV} are based on the reinterpretation of the results given by ATLAS in Ref.~\cite{Aad:2015osa,atlas-13,atlas-MSSM-13}. Specifically, we have performed a recast of an 8 TeV and 13 TeV inclusive search for a neutral $Z^\prime$ in the $\tau_{\text{had}}\tau_{\text{had}}$ channel described in Ref.~\cite{Aad:2015osa,atlas-13}. This recast sets exclusion limits on high-mass resonances in the range $0.5-2.5$~TeV but is less sensitive to resonances with masses bellow 500~GeV. In order to cover the low-mass region we performed a recast of a recent 13 TeV MSSM neutral Higgs search with $\mathcal{L}_{\text{int}}\!=\!13.2$ fb$^{-1}$ in the $\tau_{\text{had}}\tau_{\text{had}}$ channel~\cite{atlas-MSSM-13}. This last search is more sensitive to resonances in the mass range $0.2-1.2$ TeV because of better statistics due to higher luminosity. 

For the collider simulations, we have implemented the EFT and the simplified models discussed in Sec.~\ref{sec:III} with the Universal File Output (UFO) format generated by {\tt FeynRules 2}~\cite{Alloul:2013bka}. For each model we generated with {\tt Madgraph  5}~\cite{Alwall:2014hca} large samples of $pp\thinspace (b\bar b)\!\to\!\tau^+\tau^-$ events at LO. Both {\tt Pythia 6}~\cite{Sjostrand:2006za} and {\tt Pythia 8.210}~\cite{Sjostrand:2014zea} were used to decay the $\tau$-leptons, simulate parton showering and include hadronization. Any effects due to spin correlations for the $\tau$-decays were neglected. The detector response was simulated with {\tt Delphes 3}~\cite{Ovyn:2009tx} coupled with {\tt FastJet}~\cite{Cacciari:2011ma,Cacciari:2005hq} for jet clustering. The ATLAS Delphes card was modified to satisfy the object reconstruction and identification requirements used in each of the experimental searches, in particular the corresponding $\tau_{\text{had}}$-tagging and $b$-tagging efficiencies were set accordingly.


Following the $Z^\prime$ search in Ref.~\cite{Aad:2015osa,atlas-13}, events were selected if they contained at least two identified $\tau_{\text{had}}$, one with $p_T\!>\!150$~GeV~\cite{Aad:2015osa} ($p_T\!>\!110$~GeV~\cite{atlas-13}) and the other with $p_T\!>\!50$~GeV~\cite{Aad:2015osa}  ($p_T\!>\!55$~GeV~\cite{atlas-13}), no electrons with $p_T\!>\!15$~GeV, and no muons with $p_T\!>\!10$~GeV. Additionally, the visible part of the candidate $\tau_{\text{had}}\tau_{\text{had}}$ pair had to be of opposite-sign (OS) and produced back-to-back in the azimuthal plane with $\Delta\phi(\tau_1,\tau_2)\!>\!2.7$ rad. Finally, in order to reconstruct the mass of the $\tau_{\text{had}}\tau_{\text{had}}$ pair, the selected events were binned into signal regions defined by different threshold values of the total transverse mass $m_T^{\text{tot}}$ defined in Eq.~\eqref{eq:mT}.
For the recast of Ref.~\cite{Aad:2015osa} we used the $m_T^{\text{tot}}$ thresholds, observed data and expected background events from Table~4 in~\cite{Aad:2015osa}. For the recast of Ref.~\cite{atlas-13}, the thresholds $m_T^{\rm{tot}}>$ 150, 186, 231, 287, 357, 444, 551 and 684 GeV and other quantities were directly extracted from Fig.~4(f) in~\cite{atlas-13}. Our simulations and event selections were carefully validated by comparing our results with those obtained by ATLAS in~\cite{Aad:2015osa,atlas-13} for both background and signal Drell-Yan samples $pp\to \tau_{\text{had}}\tau_{\text{had}}$ mediated by $Z/\gamma^*$ in the SM and by $Z'$ in the SSM. 

The 13 TeV MSSM Higgs search~\cite{atlas-MSSM-13} uses a set of selection and kinematic cuts similar to those employed in the $Z^\prime$ searches, with the additional requirement that events be categorized according to their $b$-jet content: events with no $b$-jets belong to the $b$-veto category, while events with at least one $b$-tagged jet belongs to the $b$-tag category. Given that both categories are mutually orthogonal and use compatible binning for the final events, we decided to combine them into a fully inclusive category defined by the tighter kinematic cuts and wider $m_T^{\text{tot}}$ bins used in the $b$-tag category of~\cite{atlas-MSSM-13}. The specific selection requirements for the inclusive search recast are given by: at least one OS $\tau_{\text{had}}\tau_{\text{had}}$ pair produced back-to-back in the azimuthal plane with $\Delta\phi(\tau_1,\tau_2)\!>\!2.7$ rad, a $p_T$ requirement for the leading $\tau_{\text{had}}$ of $p_T\!>\!110$ GeV for the 2015 data set ($\mathcal{L}_{\text{int}}\!=\!3.2$ fb$^{-1}$) and $p_T\!>\!140$~GeV for the 2016 data set ($\mathcal{L}_{\text{int}}\!=\!10$ fb$^{-1}$) and a $p_T$ requirement of $p_T\!>\!65$~GeV for the sub-leading $\tau_{\text{had}}$. Following the wider binning used for the $b$-tag category in Fig.~4(e)~\cite{atlas-MSSM-13}, final events were binned into $m_T^{\text{tot}}$ intervals defined by 150, 200, 250, 300, 350, 400, 450~GeV respectively. The observed data and background events for each of these bins were extracted from Fig.~4(d) and Fig.~4(e) of~\cite{atlas-MSSM-13} and combined accordingly. We also validated our simulations and event selections by reproducing several $m_T^{\text{tot}}$ distributions in~\cite{atlas-MSSM-13} .

\end{appendix}


\begin{thebibliography}{99}        

\bibitem{Agashe:2014kda} 
  K.~A.~Olive {\it et al.} [Particle Data Group Collaboration],
  Chin.\ Phys.\ C {\bf 38}, 090001 (2014).
  doi:10.1088/1674-1137/38/9/090001
  
  \bibitem{Amhis:2014hma} 
  Y.~Amhis {\it et al.} [Heavy Flavor Averaging Group (HFAG) Collaboration],
  arXiv:1412.7515 [hep-ex] and online update at http://www.slac.stanford.edu/xorg/hfag.
  
  \bibitem{Lees:2012xj}
  J.~P.~Lees {\it et al.} [BaBar Collaboration],
  Phys.\ Rev.\ Lett.\  {\bf 109} (2012) 101802
  doi:10.1103/PhysRevLett.109.101802
  [arXiv:1205.5442 [hep-ex]].
  
\bibitem{Lees:2013uzd} 
  J.~P.~Lees {\it et al.} [BaBar Collaboration],
  Phys.\ Rev.\ D {\bf 88}, no. 7, 072012 (2013)
  doi:10.1103/PhysRevD.88.072012
  [arXiv:1303.0571 [hep-ex]].
  
\bibitem{Huschle:2015rga} 
  M.~Huschle {\it et al.} [Belle Collaboration],
  Phys.\ Rev.\ D {\bf 92}, no. 7, 072014 (2015)
  doi:10.1103/PhysRevD.92.072014
  [arXiv:1507.03233 [hep-ex]].
  
\bibitem{Abdesselam:2016cgx} 
  A.~Abdesselam {\it et al.} [Belle Collaboration],
  arXiv:1603.06711 [hep-ex].
 
      
\bibitem{Aaij:2015yra} 
  R.~Aaij {\it et al.} [LHCb Collaboration],
  Phys.\ Rev.\ Lett.\  {\bf 115}, no. 11, 111803 (2015)
  Addendum: [Phys.\ Rev.\ Lett.\  {\bf 115}, no. 15, 159901 (2015)]
  doi:10.1103/PhysRevLett.115.159901, 10.1103/PhysRevLett.115.111803
  [arXiv:1506.08614 [hep-ex]].
  
  \bibitem{Kamenik:2008tj} 
  J.~F.~Kamenik and F.~Mescia,
  Phys.\ Rev.\ D {\bf 78}, 014003 (2008)
  doi:10.1103/PhysRevD.78.014003
  [arXiv:0802.3790 [hep-ph]].
  
\bibitem{Lattice:2015rga} 
  J.~A.~Bailey {\it et al.} [MILC Collaboration],
  Phys.\ Rev.\ D {\bf 92}, no. 3, 034506 (2015)
  doi:10.1103/PhysRevD.92.034506
  [arXiv:1503.07237 [hep-lat]].
  
  \bibitem{Na:2015kha} 
  H.~Na {\it et al.} [HPQCD Collaboration],
  Phys.\ Rev.\ D {\bf 92}, no. 5, 054510 (2015)
  Erratum: [Phys.\ Rev.\ D {\bf 93}, no. 11, 119906 (2016)]
  doi:10.1103/PhysRevD.93.119906, 10.1103/PhysRevD.92.054510
  [arXiv:1505.03925 [hep-lat]].


\bibitem{Fajfer:2012vx} 
  S.~Fajfer, J.~F.~Kamenik and I.~Nisandzic,
  Phys.\ Rev.\ D {\bf 85}, 094025 (2012)
  doi:10.1103/PhysRevD.85.094025
  [arXiv:1203.2654 [hep-ph]].
  
  \bibitem{Fajfer:2012jt} 
  S.~Fajfer, J.~F.~Kamenik, I.~Nisandzic and J.~Zupan,
  Phys.\ Rev.\ Lett.\  {\bf 109}, 161801 (2012)
  doi:10.1103/PhysRevLett.109.161801
  [arXiv:1206.1872 [hep-ph]].
  
  \bibitem{Crivellin:2012ye} 
  A.~Crivellin, C.~Greub and A.~Kokulu,
  Phys.\ Rev.\ D {\bf 86}, 054014 (2012)
  doi:10.1103/PhysRevD.86.054014
  [arXiv:1206.2634 [hep-ph]].
  
  \bibitem{Celis:2012dk} 
  A.~Celis, M.~Jung, X.~Q.~Li and A.~Pich,
  JHEP {\bf 1301}, 054 (2013)
  doi:10.1007/JHEP01(2013)054
  [arXiv:1210.8443 [hep-ph]].
  
\bibitem{Greljo:2015mma} 
  A.~Greljo, G.~Isidori and D.~Marzocca,
  JHEP {\bf 1507}, 142 (2015)
  doi:10.1007/JHEP07(2015)142
  [arXiv:1506.01705 [hep-ph]].
 
\bibitem{Boucenna:2016wpr} 
  S.~M.~Boucenna, A.~Celis, J.~Fuentes-Martin, A.~Vicente and J.~Virto,
  Phys.\ Lett.\ B {\bf 760}, 214 (2016)
  doi:10.1016/j.physletb.2016.06.067
  [arXiv:1604.03088 [hep-ph]].
 
\bibitem{Dorsner:2013tla} 
  I.~Dorsner, S.~Fajfer, N.~Kosnik and I.~Nisandzic,
  JHEP {\bf 1311}, 084 (2013)
  doi:10.1007/JHEP11(2013)084
  [arXiv:1306.6493 [hep-ph]].
  
\bibitem{Sakaki:2013bfa} 
  Y.~Sakaki, M.~Tanaka, A.~Tayduganov and R.~Watanabe,
  Phys.\ Rev.\ D {\bf 88}, no. 9, 094012 (2013)
  doi:10.1103/PhysRevD.88.094012
  [arXiv:1309.0301 [hep-ph]].
  
\bibitem{Bauer:2015knc} 
  M.~Bauer and M.~Neubert,
  Phys.\ Rev.\ Lett.\  {\bf 116}, no. 14, 141802 (2016)
  doi:10.1103/PhysRevLett.116.141802
  [arXiv:1511.01900 [hep-ph]].
  
\bibitem{Barbieri:2015yvd} 
  R.~Barbieri, G.~Isidori, A.~Pattori and F.~Senia,
  Eur.\ Phys.\ J.\ C {\bf 76}, no. 2, 67 (2016)
  doi:10.1140/epjc/s10052-016-3905-3
  [arXiv:1512.01560 [hep-ph]].
  
\bibitem{Alonso:2015sja} 
  R.~Alonso, B.~Grinstein and J.~Martin Camalich,
  JHEP {\bf 1510}, 184 (2015)
  doi:10.1007/JHEP10(2015)184
  [arXiv:1505.05164 [hep-ph]].
  
\bibitem{Freytsis:2015qca} 
  M.~Freytsis, Z.~Ligeti and J.~T.~Ruderman,
  Phys.\ Rev.\ D {\bf 92}, no. 5, 054018 (2015)
  doi:10.1103/PhysRevD.92.054018
  [arXiv:1506.08896 [hep-ph]].
  
  \bibitem{LFUt}
  J.~F.~Kamenik, A.~Katz and D.~Stolarski, in preparation.
  
\bibitem{Feruglio:2016gvd} 
  F.~Feruglio, P.~Paradisi and A.~Pattori,
  arXiv:1606.00524 [hep-ph].

  \bibitem{Carrasco:2015pra} 
  N.~Carrasco {\it et al.} [ETM Collaboration],
  Phys.\ Rev.\ D {\bf 92}, no. 3, 034516 (2015),
  doi:10.1103/PhysRevD.92.034516,
  [arXiv:1505.06639 [hep-lat]].
  
\bibitem{Pappadopulo:2014qza} 
  D.~Pappadopulo, A.~Thamm, R.~Torre and A.~Wulzer,
  JHEP {\bf 1409}, 060 (2014)
  doi:10.1007/JHEP09(2014)060
  [arXiv:1402.4431 [hep-ph]].

  \bibitem{Abbiendi:2013hk} 
  G.~Abbiendi {\it et al.} [ALEPH and DELPHI and L3 and OPAL and LEP Collaborations],
  Eur.\ Phys.\ J.\ C {\bf 73}, 2463 (2013)
  doi:10.1140/epjc/s10052-013-2463-1
  [arXiv:1301.6065 [hep-ex]].


\bibitem{Buttazzo:2016kid} 
  D.~Buttazzo, A.~Greljo, G.~Isidori and D.~Marzocca,
  arXiv:1604.03940 [hep-ph].

  \bibitem{Barbieri:2006bg} 
  R.~Barbieri, L.~J.~Hall, Y.~Nomura and V.~S.~Rychkov,
  Phys.\ Rev.\ D {\bf 75}, 035007 (2007),
  doi:10.1103/PhysRevD.75.035007,
  [hep-ph/0607332].
  
  \bibitem{Baak:2014ora} 
  M.~Baak {\it et al.} [Gfitter Group Collaboration],
  Eur.\ Phys.\ J.\ C {\bf 74}, 3046 (2014),
  doi:10.1140/epjc/s10052-014-3046-5,
  [arXiv:1407.3792 [hep-ph]].

\bibitem{Fajfer:2015ycq} 
  S.~Fajfer and N.~Ko\v snik,
  Phys.\ Lett.\ B {\bf 755}, 270 (2016),
  doi:10.1016/j.physletb.2016.02.018,
  [arXiv:1511.06024 [hep-ph]].

\bibitem{Aad:2015caa} 
  G.~Aad {\it et al.} [ATLAS Collaboration],
  Eur.\ Phys.\ J.\ C {\bf 76}, no. 1, 5 (2016)
  doi:10.1140/epjc/s10052-015-3823-9
  [arXiv:1508.04735 [hep-ex]].

\bibitem{Dorsner:2014axa} 
  I.~Dorsner, S.~Fajfer and A.~Greljo,
  JHEP {\bf 1410}, 154 (2014)
  doi:10.1007/JHEP10(2014)154
  [arXiv:1406.4831 [hep-ph]].

\bibitem{Becirevic:2016yqi} 
  D.~Becirevic, S.~Fajfer, N.~Kosnik and O.~Sumensari,
  arXiv:1608.08501 [hep-ph].

  
\bibitem{CMS:2016hsa} 
  CMS Collaboration [CMS Collaboration],
  CMS-PAS-EXO-16-023.

\bibitem{Aad:2015osa} 
  G.~Aad {\it et al.} [ATLAS Collaboration],
  JHEP {\bf 1507}, 157 (2015)
  doi:10.1007/JHEP07(2015)157
  [arXiv:1502.07177 [hep-ex]].

\bibitem{atlas-13} 
  The ATLAS collaboration,
  CERN-EP-2016-164.
  [arXiv:1608.00890v2  [hep-ex]]
  
  \bibitem{Junk:1999kv} 
  T.~Junk,
  Nucl.\ Instrum.\ Meth.\ A {\bf 434}, 435 (1999)
  doi:10.1016/S0168-9002(99)00498-2
  [hep-ex/9902006].
  
\bibitem{atlas-MSSM-13} 
  The ATLAS collaboration,
  ATLAS-CONF-2016-085.

  \bibitem{Alloul:2013bka} 
  A.~Alloul, N.~D.~Christensen, C.~Degrande, C.~Duhr and B.~Fuks,
  Comput.\ Phys.\ Commun.\  {\bf 185}, 2250 (2014),
  doi:10.1016/j.cpc.2014.04.012,
  [arXiv:1310.1921 [hep-ph]].
  
  \bibitem{Alwall:2014hca} 
  J.~Alwall {\it et al.},
  JHEP {\bf 1407}, 079 (2014),
  doi:10.1007/JHEP07(2014)079,
  [arXiv:1405.0301 [hep-ph]].
  
  \bibitem{Ball:2012cx} 
  R.~D.~Ball {\it et al.},
  Nucl.\ Phys.\ B {\bf 867}, 244 (2013)
  doi:10.1016/j.nuclphysb.2012.10.003
  [arXiv:1207.1303 [hep-ph]].
  
    
  
  \bibitem{Ball:2014uwa} 
  R.~D.~Ball {\it et al.} [NNPDF Collaboration],
  JHEP {\bf 1504}, 040 (2015)
  doi:10.1007/JHEP04(2015)040
  [arXiv:1410.8849 [hep-ph]].
  
  \bibitem{Lim:2016wjo} 
  M.~Lim, F.~Maltoni, G.~Ridolfi and M.~Ubiali,
  arXiv:1605.09411 [hep-ph].

  \bibitem{Dittmaier:2011ti} 
  S.~Dittmaier {\it et al.} [LHC Higgs Cross Section Working Group Collaboration],
  doi:10.5170/CERN-2011-002,
  arXiv:1101.0593 [hep-ph], and online updates.
  
    \bibitem{Drees:1990dq} 
  M.~Drees and K.~i.~Hikasa,
  Phys.\ Lett.\ B {\bf 240}, 455 (1990)
  Erratum: [Phys.\ Lett.\ B {\bf 262}, 497 (1991)].
  doi:10.1016/0370-2693(90)91130-4

\bibitem{Khachatryan:2016vau} 
  G.~Aad {\it et al.} [ATLAS and CMS Collaborations],
  JHEP {\bf 1608}, 045 (2016)
  doi:10.1007/JHEP08(2016)045
  [arXiv:1606.02266 [hep-ex]].
  
  \bibitem{Efrati:2016uuy} 
  A.~Efrati, J.~F.~Kamenik and Y.~Nir,
  arXiv:1606.07082 [hep-ph].

  
\bibitem{Dumont:2016xpj} 
  B.~Dumont, K.~Nishiwaki and R.~Watanabe,
  Phys.\ Rev.\ D {\bf 94}, no. 3, 034001 (2016),
  doi:10.1103/PhysRevD.94.034001,
  [arXiv:1603.05248 [hep-ph]].
  
  
\bibitem{Khachatryan:2015bsa} 
  V.~Khachatryan {\it et al.} [CMS Collaboration],
  JHEP {\bf 1507}, 042 (2015),
  doi:10.1007/JHEP07(2015)042,
  [arXiv:1503.09049 [hep-ex]].

  
  
  

  
  \bibitem{Andrea:2011ws} 
  J.~Andrea, B.~Fuks and F.~Maltoni,
  Phys.\ Rev.\ D {\bf 84}, 074025 (2011)
  doi:10.1103/PhysRevD.84.074025
  [arXiv:1106.6199 [hep-ph]];
  J.~F.~Kamenik and J.~Zupan,
  Phys.\ Rev.\ D {\bf 84}, 111502 (2011)
  doi:10.1103/PhysRevD.84.111502
  [arXiv:1107.0623 [hep-ph]];
  E.~Alvarez, E.~Coluccio Leskow, J.~Drobnak and J.~F.~Kamenik,
  Phys.\ Rev.\ D {\bf 89}, no. 1, 014016 (2014)
  doi:10.1103/PhysRevD.89.014016
  [arXiv:1310.7600 [hep-ph]].
  
\bibitem{Sjostrand:2006za}
 T. Sj\"ostrand, S. Mrenna and P. Z. Skands,
      JHEP {\bf 0605}, 026 (2016)
  
  
  \bibitem{Sjostrand:2014zea}
      T. Sj\"ostrand {\it et al.},
      Comput. Phys. Commun. {\bf 191} 159-177 (2015),
      [arXiv:1410.3012 [hep-ph]].
  
  \bibitem{Ovyn:2009tx}
      S. Ovyn, X. Rouby and V. Lemaitre,
      CP3-09-01
      [arXiv:0903.2225 [hep-ph]],
      archivePrefix  = "arXiv",
  
 \bibitem{Cacciari:2011ma}
  M.Cacciari, G. P. Salam, G. Soyez,
  Eur. Phys. J. {\bf C72} 1896 (2012),
 doi:10.1140/epjc/s10052-012-1896-2,
 [arXiv:1111.6097 [hep-ph]].
 
 \bibitem{Cacciari:2005hq}
     M. Cacciari and G. P. Salam,
      Phys. Lett. {\bf B641} 57-61(2006),
      [arXiv:hep-ph/0512210 [hep-ph]]
 
 
  
  

  
  

\end{thebibliography}
\end{document}